\documentclass[12pt,preprint,notoc]{JHEP3}

\usepackage{epsfig,multicol}
\usepackage{amssymb}
\usepackage{graphics}
\usepackage{amsmath,latexsym}
\def\beq{\begin{eqnarray}}
\def\eeq{\end{eqnarray}}

\usepackage{amssymb}
\usepackage{amsmath}
\usepackage{graphicx}
\def\beq{\begin{eqnarray}}
\def\eeq{\end{eqnarray}}
\def\ln{\,\mbox{ln}\,}

\def\La{\Lambda}

\newcommand{\rD}{\rho_D}
\newcommand{\OM}{\Omega_M}

\newcommand{\OMo}{\Omega_{M}^0}
\newcommand{\ORo}{\Omega_{R}^0}
\newcommand{\OL}{\Omega_{\Lambda}}

\newcommand{\OLo}{\Omega_{\Lambda}^0}
\newcommand{\OX}{\Omega_{X}}
\newcommand{\OXo}{\Omega_{X}^0}

\newcommand{\OD}{\Omega_{D}}
\newcommand{\ODo}{\Omega_{D}^0}

\newcommand{\rc}{\rho_c}
\newcommand{\rco}{\rho_{c}^0}

\newcommand{\rM}{\rho_M}

\newcommand{\rX}{\rho_X}
\newcommand{\pX}{p_X}
\newcommand{\wX}{w_X}

\newcommand{\wR}{w_R}

\newcommand{\rL}{\rho_{\CC}}
\newcommand{\pL}{p_{\CC}}

\newcommand{\pD}{p_D}

\newcommand{\wL}{w_{\CC}}
\newcommand{\CC}{\Lambda}

\newcommand{\we}{w_{e}}

\newcommand{\tOM}{\tilde{\Omega}_M}

\newcommand{\tOD}{\tilde{\Omega}_{D}}

\newcommand{\G}{{\cal G}}

\def\La{\Lambda}

\title{Effective growth of matter density fluctuations in the running
$\Lambda$CDM and $\Lambda$XCDM models}
\author{\hspace{0.3cm}
Javier Grande $^{1}$, Reuven Opher $^{2}$, Ana Pelinson $^{1}$, Joan Sol\`a $^{1,3}$\\
$^{1}\,$ High Energy Physics Group, Dept. ECM,  Universitat de Barcelona,\\
\hspace{0.2cm}
Diagonal 647, 08028 Barcelona, Catalonia, Spain\\
$^{2}\,$ IAG, Universidade de
S\~{a}o Paulo, Rua do Mat\~{a}o 1226,\\
\hspace{0.2cm} Cidade Universit\'aria, CEP 05508-900, S\~{a}o Paulo,
SP, Brazil\\
$^{3}\,$ Institut de Ci\`{e}ncies del Cosmos, UB, Barcelona.\\
\hspace{0.2cm} E-mails: jgrande@ecm.ub.es, opher@astro.iag.usp.br,
apelinson@ecm.ub.es, sola@ifae.es\\
}

\preprint{UB-ECM-PF 07/26}

\abstract{We investigate the matter density fluctuations
$\delta\rho_M/\rho_M$ for two dark energy (DE) models in the
literature in which the cosmological term $\La$ is a running
parameter. In the first model, the running $\La$CDM model, matter
and DE exchange energy, whereas in the second model, the $\La$XCDM
model, the total DE and matter components are conserved separately.
The $\CC$XCDM model was proposed as an interesting solution to the
cosmic coincidence problem. It includes an extra dynamical
component, the ``cosmon'' $X$, which interacts with the running
$\Lambda$, but not with matter. In our analysis we make use of the
current value of the linear bias parameter, $b^2(0)= P_{GG}/P_{MM}$,
where $P_{MM}\propto (\delta\rho_M/\rho_M)^2$ is the present matter
power spectrum and $P_{GG}$ is the galaxy fluctuation power
spectrum. The former can be computed within a given model, and the
latter is found from the observed LSS data (at small $z$) obtained
by the 2dF galaxy redshift survey. It is found that
$b_{\Lambda}^2(z\simeq 0)=1$ within a $10\%$ accuracy for the
standard $\La$CDM model. Adopting this limit for any DE model and
using a method based on the effective equation of state for the DE,
we can set a limit on the growth of matter density perturbations for
the running $\La$CDM model, the solution of which is known. This
provides a good test of the procedure, which we then apply to the
$\CC$XCDM model in order to determine the physical region of
parameter space, compatible with the LSS data. In this region, the
$\CC$XCDM model is consistent with known observations and provides
at the same time a viable solution to the cosmic coincidence
problem.}

\begin{document}

\section{Introduction\label{sec1}}

For the last two decades or so, the  $90$-years-old history of the
cosmological constant problem\,\cite{weinRMP} has turned into the
history of the dark energy (DE) problem\,\cite{CCRev}. A massive
attempt has been underway in recent times to supersede the
cosmological term $\CC$ in Einstein's equations by a variety of
different entities which come under the mysterious name of DE, i.e.,
a purported new substance or, for that matter, an effective cause
that is responsible for the observed accelerated expansion of the
Universe\,\cite{Supernovae}. In particular, there is the popular
idea that the DE is associated with a cosmological dynamical scalar
field (quintessence and the like)\,\cite{quintessence,DEquint},
which fully supplants the preeminent role played by  $\CC$ for a
long time in different chapters of modern
cosmology\,\cite{weinRMP,CCRev}. Within this broader context, the
cosmological constant (CC) has been relegated to a back seat in the
cosmological scenario or, at least, has been degraded to represent
just one among a host of possibilities, proposed to explain the
nature of the DE. In spite of this situation, the $\CC$ term and the
standard model ($\CC$CDM model) of modern cosmology have been
thriving rather well and have survived, essentially unscathed, the
entire set of observational tests to which they have been exposed up
to the present time (see \,\cite{WMAP3Y} for a summary of the
experimental situation).

This state of affairs somehow suggests that, rather than trying to
completely get rid of the CC term and replace it by some Ersatz
entity, perhaps it would be a better idea to keep it and try to
explain some of the unsatisfactory features of the cosmological
standard model in terms of possible, unsuspected dynamical features
of $\CC$ and/or by introducing other dynamical complements to it.
For example, while it is very hard to accept a small and strictly
constant value of $\CC$ throughout the entire history of the
Universe, a slowly evolving DE looks more promising. Actually, this
potentially dynamical character of the DE is the main motivation for
introducing quintessence-like ideas\,\cite{DEquint}. However, the
contribution from the vacuum energy, most likely represented by the
$\CC$ term, is still there and remains a good candidate to be
considered. Therefore, instead of exchanging it for a dynamical new
object, it seems more economical to just admit that $\CC$ may hide
some small evolution (``running'') with time or energy. Even more
ambitiously, one may entertain the possibility that the CC could be
a running parameter in quantum field theory (QFT) in curved
space-time, i.e., an effective charge in the sense of the
renormalization group (RG). We may call this scenario the ``running
$\CC$CDM model''. An interesting proposal along these lines was
developed in \,\cite{JHEPCC1,IRGA03,SSS1} (for a review, see e.g.,
\cite{IRGAC06}), and was put to the test in \cite{RGTypeIa,SS1,SS2}
(see also \cite{Babic,Betivegna,Guberina06}).

A particularly acute cosmological problem, that could be alleviated
by introducing a dynamical DE, is known as the  ``coincidence
problem''\,\cite{Steinhardt}, i.e., why do the matter and DE
densities happen to be of the same order of magnitude, precisely in
the current Universe. For example, why is the latter not one
thousand or one million times smaller at present?  The popular idea
of a dynamical scalar field replacing the cosmological constant
$\CC$ was largely motivated by the possibility of having a framework
where one could try to solve this conundrum. Another option for
tackling this problem is to start with the running $\CC$CDM model,
which does not deviate much from the standard cosmological model,
and add to it a dynamical entity, $X=X(t)$, which was called the
``cosmon''\,\cite{LXCDM1}\,\footnote{See \cite{LXCDM2,GSSIRGAC12}
for alternative cosmological models involving the cosmon and both
variable cosmological term $\CC$ and Newton's coupling $G$.}. In
contrast to the standard quintessence point of view (where $\CC$CDM
is replaced by $X$CDM), here the $\CC$ term is not substituted by a
scalar field. Instead, it is assumed that there may be multiple
sources of DE (perhaps some fundamental and others effective),
including the vacuum energy, which is tied to $\CC$. The other may
be collectively represented by the effective entity $X$. In this
model, which was called the $\CC$XCDM model in\,\cite{LXCDM1},
matter and the total DE are conserved separately. However, the DE
density here is not just $\rL=\CC/(8\pi\,G)$, but is the sum of
$\rL$ and the cosmon density, $\rX$, i.e., $\rD=\rL+\rX$. This total
DE is locally and covariantly conserved with the expansion of the
Universe. {The advantage of upgrading the $\CC$CDM model into the
$\CC$XCDM is that it allows for the possibility of dynamical
interplay between $X$ and $\CC$ within a scenario where the total
matter and DE densities are individually conserved}\,\footnote{The
name \textit{cosmon} was first introduced in\,\cite{PSW}. Here we
use it in a generalized sense for any additional component(s) of the
DE, other than $\CC$, provided that the total DE density remains
covariantly conserved.}. This interplay is essential in order to
provide a solution for the coincidence problem as well as to allow
for the $\CC$XCDM model to mimic the standard $\CC$CDM model at the
present time. Indeed, for a wide range of cosmological redshifts
(including the full span accessible to supernovae data), the
effective DE pressure and density $(\pD,\rD)$ in the $\CC$XCDM model
may simulate a constant $\CC$ behavior, $\pD\simeq -\rD$, to an
arbitrary high degree of approximation\,\cite{LXCDM1}.

It should be emphasized that, in contrast to the quintessence point
of view, the  entity $X$ in the $\CC$XCDM model need not be a scalar
field. In fact, no physical substratum (e.g. a physical fluid) is
assumed behind it. The essential condition defining $X$ is the local
covariant conservation of the total DE, $\rD=\rL+\rX$, independent
of matter. In particular, $X$ could be the effective behavior of a
more complete theory, comprising Einstein's gravity as a particular
case. For example, it could represent the net behavior of higher
order terms in the effective action.

An essential aspect to address in any proposed model for the DE is
its ability to reproduce the existing observations. The $\CC$XCDM
model has been already put to the test concerning nucleosynthesis
bounds and the equation of state (EOS) behavior at small and large
redhifts, that are relevant to supernovae and CMB
data\,\cite{LXCDM1}. However, one of the most important tests yet to
be made is the verification of the existence of a region of
parameter space that is also compatible with the data on the large
scale structure (LSS) formation. This is, in fact, the main aim of
the present paper. The LSS is contained in the galaxy fluctuation
power spectrum $P_{GG}$ which has been measured by the 2 degree
Field Galaxy Redshift Survey (2dFGRS), for example, \cite{Cole05}.
It is remarkable that the joint 2dFGRS and CMB analysis presented
in\,\cite{Lahav02} shows that there is a good agreement of the LSS
data with the measurements of the CMB anisotropies for the $\CC$CDM
model as well as with numerical simulations of galaxy
formation\,\cite{somerville01}.
\\
\indent On the theoretical side, this data must be reproduced by the
predicted matter power spectrum, $P_{MM}$, for any successful model
of structure formation. Therefore, for very large scales, it is to
be expected that the linear bias parameter,  $b^2\equiv
P_{GG}/P_{MM}$\,\cite{Lahav02}, should behave as a definite,
scale-independent, quantity at small cosmological redshifts, i.e.,
when the distribution of galaxies had enough time to be correlated
with the mass distribution in the Universe. This is, in fact, what
is obtained in the successful standard $\CC$CDM model, where the
observed present value of $b^2$ turns out to be $1$ within a $10\%$
accuracy\,\cite{Cole05,Lahav02}. It seems reasonable to extend this
criteria to any other cosmological model aiming at a good
description of the presently observed LSS. For example, in the power
spectrum $P_{MM}$ for the running $\CC$CDM
model\,\cite{JHEPCC1,IRGA03}, which has been fully studied in
\cite{FSS1}, the matter fluctuations have been solved in a framework
where they are coupled with the perturbations in the DE, in this
case represented by the running $\CC$, described by the parameter
$\nu$. Its comparison with the galaxy fluctuation power spectrum
$P_{GG}$\,\cite{Cole05} puts stringent limits on the fundamental
parameter $\nu$ of this model. A non-vanishing value of $\nu$
produces a time evolution of $\CC$. The explicit solution of the
model shows that values $|\nu|>10^{-4}$ are most likely
excluded\,\cite{FSS1}.
\\
\indent  Here, we wish to address the LSS test for the $\CC$XCDM
model. The test includes three parameters: $\nu$ (the time evolution
of $\CC$), $\wX$ (the effective barotropic index of the cosmon), and
$\OLo$ (the current CC density in units of the critical density). In
this model as well, a non-vanishing value of $\nu$ entails an
evolving $\CC$, but in contradistinction to the running $\CC$CDM
model or the interactive quintessence models\,\cite{interactiveQ},
there is no crosstalk between matter and DE, which is an attractive
feature. The exact analysis of matter density perturbations is more
complicated in the $\CC$XCDM model, and we tackle it using the
following two-step ``effective method''. First of all, we note that
it is possible to ascribe an effective EOS to a variable $\CC$
model. This has been explored in detail in \cite{SS1,SS2}. Using
these results, we apply the effective EOS approach to the matter
perturbation equations following\,
\cite{Ferreira97,Ma,LinderJenkins03}. This enables us to obtain an
approximate treatment of the growth of matter perturbations, in
which the DE perturbations are neglected and all the DE effects are
encoded in the effective EOS, $\pD=\we\,\rD$ and in the ratio of DE
to matter densities, $r=\rD/\rM$. Secondly, to obtain useful bounds
on the parameters of the model, we use the linear bias parameter,
$b^2(z)= P_{GG}/P_{MM}$. Specifically, we impose the condition
(``F-test''\,\cite{Ftest,vdecay}) that its value cannot deviate from
the $\CC$CDM value by more than $10\%$ at $z=0$. From the above, the
F-test should be essentially equivalent (although not identical) to
requiring that $b^2(0)=1\pm 0.1$\,\cite{Cole05,Lahav02}. Some
concrete applications of this test can be found in\,\cite{Ftest}. In
the present paper we look for the viable physical region of
parameter space for the $\CC$XCDM model, using the aforementioned
effective method. However, to check its efficiency when applied to
non-trivial models with variable $\CC$, we first address the running
$\CC$CDM model. {It is important to emphasize that, in the running
$\CC$CDM model, the DE (represented by the variable CC) and matter
fluids exchange energy and, therefore, are interacting components of
the cosmic medium. Fortunately, since we know the results of a
full-fledged treatment of density perturbations (of both matter and
DE) in this model\,\cite{FSS1}, we can compare them with those
obtained from the effective approach, in which DE and matter are
treated as conserved, non-interacting components and the
perturbations of the DE are neglected. This effective approach is
meaningful, provided we arrange that the expansion history is the
same as that of the original running $\CC$CDM model. We may call
this alternative representation the ``DE picture''\,\cite{SS2}
because it is close to the standard quintessence representation of
the DE\,\cite{DEquint}.} The comparison provides an excellent test
for the effective method used in the DE picture. Finally, we apply
the effective EOS procedure, in combination with the F-test, to the
more complicated situation of the $\CC$XCDM model and obtain the
corresponding region of parameter space that is compatible with the
LSS data. The existence of this region strengthens, once more, the
viability and likelihood of the $\CC$XCDM model as a robust solution
of the coincidence problem\,\cite{LXCDM1,LXCDM2,GSSIRGAC12}.

The structure of the paper is as follows. In the next section, we
review the effective EOS approach to the computation of matter
density perturbations and discuss the F-test. In section
\ref{sect:runningLCDM}, we apply the effective EOS approach and the
F-test to the running $\CC$CDM model. In sections \ref{sect:LXCDM}
and \ref{sect:numerical} we describe the $\CC$XCDM model and present
a detailed numerical analysis of the matter density perturbations in
order to determine the physical region of parameter space. The final
section is devoted to discussion and conclusions.

\section{Dark energy, density fluctuations and the F-test\label{sect:growth}}

In this section, we discuss the effective approach to the
computation of the linear matter density fluctuations for dark
energy (DE) models in which the matter components are canonically
conserved\,\cite{Ferreira97,Ma,LinderJenkins03}, as well as the
definition of the F-test\,\cite{Ftest}. The main aim is to lay the
groundwork for section \ref{sect:runningLCDM}, where we apply them
to a cosmological model with a variable cosmological term $\CC$, for
which an effective equation of state (EOS) can be
defined\,\cite{SS1,SS2}. For a general DE model within the flat
Universe, the Friedmann equation can be written in terms of the
normalized matter and DE densities as
\begin{equation}
\frac{H^2(a)}{H_0^2} = \Omega_{M}(a) +
\Omega_{D}(a)\,,\,\,\,\,\,\,\,\,\,\
\label{friedeq}
\end{equation}
where $H_0$ is the present value for the Hubble parameter and
$a(z)=1/(1+z)$ is the scale factor in terms of the cosmological
redshift $z$. The matter and DE densities are normalized in terms of
the present critical density $\rho^0_c\equiv 3H^2_0/8\pi G$:
\begin{equation}\label{Omegas1}
\Omega_{M}(a)\equiv
\frac{\rho_M(a)}{\rho^0_c}\,,\,\,\,\,\,\,\,\,\,\,\ \
\Omega_{D}(a)\equiv \frac{\rho_D(a)}{\rho^0_c}\,.
\end{equation}
It will prove useful to also define the set of ``instantaneous''
cosmological parameters at cosmic time $t$, when the scale factor
was $a=a(t)$,
\begin{equation}\label{Omegas2}
\tOM(a)=\frac{\rM(a)}{\rc(a)}\,,\ \ \ \ \ \ \
\tOD(a)=\frac{\rD(a)}{\rc(a)}\,,
\end{equation}
where $\rc(a)=3H^2(a)/8\pi\,G$ is the critical density at the same
instant of cosmic time $t$. These parameters should not be confused
with those in (\ref{Omegas1}). We will use both sets, depending on
the situation, and for this reason, the parameters in
(\ref{Omegas2}) are written with a tilde. Notice that only these
parameters satisfy the normalized cosmic sum rule at any time:
\begin{equation}\label{sumrule1}
\tOM(a)+\tOD(a)=1\,.
\end{equation}
While the parameters of the original set (\ref{Omegas1}) also add up
to one at $t=t_0$ (present time), they satisfy the relation
(\ref{friedeq}), in general.

If matter and DE are individually conserved (``self-conserved''), we
can split the overall conservation law into two equations,
\beq {\rho}^{\prime}_M(a)\,+\,\frac{3}{a} \rho_M(a)\,=\,0\,,
\label{conslawmm} \eeq
\beq {\rho}^{\prime}_D(a)\,+\,\frac{3}{a} (1+\we)\,\rD(a)\,=\,0\,,
\label{conslawde}\eeq
where the primes indicate differentiation with respect to the scale
factor, $f'\equiv d\,f/da$. In the equations above we have
parameterized the relation between the pressure and density of the
dark energy and matter as follows:
\beq p_D=\we\rho_D\,, \qquad p_M=0\,, \eeq
where $\we$ is the effective equation of state (EOS) ``parameter''
of the DE. It is not, in general, a constant parameter, but a
function of the scale factor $a$, namely $\we\equiv \we (a)$.
Therefore, we do not expect the DE to behave as a simple barotropic
fluid. The solution of (\ref{conslawde}) can be expressed in terms
of the normalized DE density defined in (\ref{Omegas1}):
\begin{equation}\label{solOMD}
 \Omega_{D}(a)=\Omega_{D}^0\,\exp{\left\{-3\,\int_1^a\frac{da'}{a'}\,[1+\we(a')]\right\}}\,,
\end{equation}
where $\Omega_{D}^0=\Omega_{D}(a=1)$ is the normalized DE density at
present. Therefore, the EOS coefficient is obtained from
\begin{equation}
\we(a)= -1-\frac{a}{3}
\frac{1}{\Omega_{D}(a)}\frac{d{\Omega_{D}(a)}}{da}\,.\label{effw}
\end{equation}
Notice that we consider only non-relativistic, pressureless, matter
because our perturbation calculation refers only to the epoch of
structure formation. Since the matter conservation law is decoupled,
the normalized matter density following from (\ref{conslawmm}) reads
\begin{equation}\label{rho}
\OM(a)={\Omega^0_M}a^{-3}\,,
\end{equation}
where $\Omega_M^0$ is the present normalized total matter density.

Our analysis of matter density perturbations applies after the
radiation dominated era.  Thus, we take $z$ (alternatively $a$) from
about the recombination epoch $z_{\rm rec} \simeq 1100$ (i.e
$a\simeq 10^{-3}$) to $z = 0$ ($a=1$, today). Moreover, it applies
only to sufficiently large scales, where the perturbations follow
the linear regime, as we discuss below. On scales within the
horizon, fluctuations in the dark energy density disperse
relativistically and the DE component becomes smooth, i.e., the
density perturbations $\delta\rD$ can be considered negligible. In
these conditions, the evolution of the matter density fluctuations
$\delta\rM$ with the cosmic time $t$ can be computed from the
well-known equation of motion in the Newtonian
approximation\,\cite{cosmobooks},
\begin{equation}
\ddot{\delta}_M+2\,H\,\dot{\delta}_M-4\pi\,G\,\rM\,\delta_M =0\,,
 \label{ggrowth1}
\end{equation}
where, again, pressure effects (in this case pressure perturbations)
are neglected in the matter dominated era. Here $G$ is the Newton
constant, $\delta_M\equiv\delta\rho_M/\rho_M$ is the fractional
matter density perturbation (density contrast) and
$\dot{\delta}_M\equiv d\,\delta_M/dt$. Within this framework,
$\delta_M$ is the  linear growth of density fluctuations.

The previous equation is also valid in the presence of the DE,
provided that $\delta\rD$ is indeed negligible, as assumed. In this
case, $H$ is, of course, affected by the corresponding smooth
background density contribution from $\rD$. By the same token the
above equation is also valid in the presence of the spatial
curvature term, $K/a^2$, since it can also be treated as a smooth
source (similarly to the DE). Although we will restrict ourselves to
the flat space case, the following reformulation of (\ref{ggrowth1})
is valid for any spatial curvature. Let us write the general
Friedmann's equation as
\begin{equation}\label{Friedmann}
H^2(a)=\frac{8\,\pi\,G}{3}\left[\rM(a)+\rD(a)-\frac{3K}{8\pi G
a^2}\right]=\frac{8\,\pi\,G}{3}\,\rc(a)=\frac{8\pi\,G}{3}\,\frac{\rM(a)}{\tOM(a)}\,,
\end{equation}
where $\tOM(a)$ was defined in (\ref{Omegas2}). It is then easy to
see that Eq.\,(\ref{ggrowth1}) can be written as
\begin{equation}
\ddot{\delta}_M+2\,H\,\dot{\delta}_M-\frac32\,H^2\,\tOM(a)\,\delta_M=0\,.
 \label{ggrowth2}
\end{equation}
We wish to solve this equation for some non-trivial scenarios.
First, we introduce some cosmetic changes in (\ref{ggrowth2}), which
prove very useful for practical purposes\,\cite{LinderJenkins03}, as
they allow us to apply this method to variable $\CC$
models\,\cite{SS1,SS2}.

We start from (\ref{ggrowth2}), for $\delta_M$, which, interestingly
enough, can be conveniently recast such that the effective equation
of state (EOS) of the DE in the given model appears explicitly. We
first trade the derivative with respect to the cosmic time for the
derivative with respect to the cosmic factor:
$\dot{\delta}_M=a\,H\,\delta_M'$, where $\delta_M'\equiv
d\,\delta_M/da$. Similarly,
$\ddot{\delta}_M=(a\,H^2+a\,\dot{H})\,\delta_M'+(a\,H)^2\,\delta_M''$.
We observe that we can eliminate $\dot{H}$ (i.e. the time variation
of the Hubble function) from the last equation by noting that, in
the matter dominated epoch, it can conveniently be written as
(henceforth, we confine ourselves to the flat case only)
\begin{equation}\label{Hdot}
\dot{H}=-4\,\pi\,G\,\left[\rM+(1+\we)\,\rD\right]=-\frac32\,H^2\,\left[1+\frac{\we\,r}{1+r}\right]\,,
\end{equation}
where $\we=\pD/\rD$ is the aforementioned effective EOS parameter of
the DE and $r=r(a)$ is the ratio between the DE and matter densities
at any given time\,\cite{LXCDM1}:
\begin{equation}\label{ra}
r(a)=\frac{\Omega_D(a)}{\Omega_M(a)}=\frac{\tOD(a)}{\tOM(a)}=\frac{\rD(a)}{\rM(a)}\,.
\end{equation}
The relation (\ref{ra}) will play a relevant role in the study of
the cosmic coincidence problem within the $\CC$XCDM model (see
section \ref{sect:LXCDM}). Substituting the previous equations into
Eq.\,(\ref{ggrowth2}), we finally obtain
\begin{equation}
\delta_M''(a)+ \frac{3}{2}\left[1-\frac{{\we (a)}\,r(a)}{1+r(a)}
\right]\frac{\delta_M'(a)}{a}- \frac{3}{2}\,\frac{1}{1+r(a)}\
\frac{\delta_M(a)}{a^2}=0\,. \label{ggrowth3}
\end{equation}
This will be our master equation to evaluate the ``effective''
growth of linear density fluctuations. As we said above, we assume
that the DE component becomes smooth within the horizon. The DE
effects enter our calculations only through the effective EOS
function (\ref{effw}) and the ratio (\ref{ra}), and, thus, the
linear growth of matter fluctuations is computed in an effective
way. We also assume that the DE density was negligible at the
recombination era and remained so until $z\lesssim 10$, when all
relevant modes for LSS formation had already entered the horizon. In
particular, the perturbation amplitude at recombination, when the
CMB was formed, is independent of the particular DE model since
$\rD\ll\rM$ at that epoch.

To better assess the meaning of Eq.\,(\ref{ggrowth3}), let us first
consider two simple examples. In the absence of DE ($\tOD=0$ and,
thus, $r=0$) the growing mode solution of the above equation in the
matter dominated epoch ($a\sim t^{2/3}$) is very simple and
well-known, $\delta_M\sim a$. However, in the presence of DE, the
growing mode solution is more complicated. Assuming a time interval
not very large such that $\tOD$ and $\OD$ remain approximately
constant, we can take $\we=-1$ and, hence, (\ref{ggrowth3}) reads
\begin{equation}
\delta_M''(a)+ \frac{3}{2}(1+\tOD)\frac{\delta_M'(a)}{a}-
\frac{3}{2}(1-\tOD)\frac{\delta_M(a)}{a^2}=0\,. \label{ggrowth3ex}
\end{equation}
Looking for a power-law solution, $\delta\sim a^p$, in the limit
$\tOD\ll 1$, we find
\begin{equation}\label{deltaOD}
\delta_M\sim a^{1-6\tOD/5}\sim a\,(1-\frac65\,\tOD\,\ln a)\,.
\end{equation}
This equation conveys, very clearly, the physical idea of growth
suppression when a (positive) DE density is present within the
horizon. Although $\tOD$ is not constant in general, this
qualitative feature should persist in more realistic situations.
Furthermore, if the DE has some smooth dynamical behavior, we cannot
exclude that $\tOD$ could have been negative for some period in the
past. If so, the previous equation also shows that structure
formation was reinforced during that period. In this paper, we wish
to study the exact numerical solutions of (\ref{ggrowth3}) in some
non-trivial scenarios, where the cosmological term is not only
arbitrary and non-vanishing, but is evolving smoothly with time.
Specifically, we wish to solve (\ref{ggrowth3}) for both the running
$\CC$CDM and $\CC$XCDM models, mentioned in the introduction. The
clue to doing this is to mimic these variable $\CC$ models through a
non-trivial effective EOS, $\we=\we(a)$.

From the above discussion, it follows that the linear growth should
essentially behave as $\delta_M=a$ near $z=z_{\rm rec}$. We are
going to use this property as an exact boundary condition for
solving (\ref{ggrowth3}), together with $\delta_M'=1$ at
recombination. Equivalently, if we introduce the standard linear
growth suppression factor, $\G(a)=\delta_M(a)/a$,
Eq.\,(\ref{ggrowth3}) is readily seen to transform into
\begin{equation}
\G^{\prime\prime}(a)+
\left[\frac{7}{2}-\frac{3}{2}\frac{\we(a)\,r(a)}{1+r(a)}
\right]\frac{\G^{\prime}(a)}{a}+
\frac{3}{2}\frac{[1-\we(a)]\,r(a)}{1+r(a)}\ \frac{\G(a)}{a^2}=0\,,
\label{ggrowth4}
\end{equation}
which we solve with the boundary conditions $\G(z_{\rm rec})=1$ and
$\G'(z_{\rm rec})=0$.

To compare our predictions with observations, it is useful to invoke
the linear bias factor. Its present value is defined as $b^2(0)=
P_{GG}/P_{MM}$, where $P_{MM}$ is the  computed matter power
spectrum at the present epoch, starting from the $P_{MM}$ at the
recombination epoch (obtained from the observed CMB) and $P_{GG}$ is
the galaxy distribution power spectrum (obtained from the observed
galaxy-galaxy correlation function). Therefore, $b^2(0)$ measures
the difference in clustering between galaxies and mass fluctuations,
i.e., it parameterizes the degree by which light traces mass. It can
be related to the rms mass fluctuations on random spheres of radius
$R\,h^{-1}$ Mpc, typically with $R=8$, and hence connected with
$\sigma_8$ \,\cite{cosmobooks}. In general, the bias factor can also
be defined in the non-linear regime. Here, however, we are concerned
only with the \textit{linear} bias factor\,\cite{Lahav02}, which
measures the difference in clustering between galaxies and mass
fluctuations at very large scales, namely at scales for which the
wave-numbers of the Fourier modes are in the range $0.02<k
<0.15\,h\,$Mpc$^{-1}$. The observational data concerning the linear
regime do, in fact, lie in this range\,\cite{Cole05}. Expressing the
scales in terms of the Hubble radius $H_0^{-1}\simeq
3000\,h^{-1}\,$Mpc ($h\simeq 0.7$), this implies values of $k$ up to
$450\,{H_0}$. Thus, the minimum length scale explored by the
available LSS data is of order of $10$ Mpc $\sim 8\,h^{-1}$ Mpc
spheres.

For the computation of the linear bias factor, we need the matter
power spectrum, whose general structure is
\begin{equation}\label{Powers}
 P_{MM}(k,a) \propto
\left(\frac{\delta\rM}{\rM}\right)^2=\delta_M^2(k,a)=a^2\,\G^2(k,a)\,,
\end{equation}
where $\delta_M(k,a)$ and $\G(k,a)$ are the Fourier transforms of
the corresponding quantities. In general, the correlation function
for the mass distribution need not coincide with the correlation
function for galaxies. Rather, a ``bias'' between the two is
expected\,\cite{cosmobooks}. However, at the LSS level, it is also
expected that the value of the linear bias should be some
scale-independent number in the late epochs of structure formation
($z\simeq 0$), namely when galaxies have had time enough to be
correlated with the mass, or equivalently, when the gravitational
pull has drifted them to overdense regions. As a matter of fact, the
observed galaxy power spectrum $P_{GG}$, emerging from the final
2dFGRS catalog, did indicate these features very clearly. Most
remarkably, the data pointed to the value $b_{\Lambda}^2(z\simeq 0)=
1.0$ to within a $10\%$ accuracy for the $\Lambda$CDM model
\cite{Cole05}, i.e., when $ P_{MM}$ in (\ref{Powers}) is computed
from the growth factor for the standard cosmological model,
characterized by strictly constant $\CC$. This is in agreement with
the previous result, $b(L_S,z\simeq 0)=1.10\pm 0.08$, of the 2dFGRS
collaboration for the APM-selected massive galaxies ($L_S=1.9
L_{\ast}$), averaged over all types\,\cite{Lahav02}, indicating that
there is one $L_{\ast}$ galaxy per dark matter halo of mass $\sim
10^{13}\,h^{-1}\,M_{\odot}$ at the present epoch.

As indicated before, in the evaluation of the bias factor, $P_{GG}$
is fixed by the LSS data, whereas $P_{MM}$ is a theoretical quantity
-- and, hence, model dependent. However, the very good prediction
(at the $10\%$ level) of the bias factor by the $\CC$CDM near
$b^2(0)=1$, suggests the following strategy to put limits on new
models of structure formation. Rather than computing $P_{MM}$ in
detail for the given model, it can just be compared to the $\CC$CDM
model. This is, of course, simpler than the full computation of
$P_{MM}$, in which details of the transfer function and other
normalization factors must be included in the prefactor of
(\ref{Powers}). These prefactors are common and cancel in the ratio.
Therefore, in the present analysis, we adhere to the following
``F-test''(first proposed in\,\cite{Ftest}), which uses the method
of comparison to evaluate the viability of a given DE model. To
gauge the deviation of the power spectrum of the model, $P_{MM}(a)$,
with respect to that of the $\CC$CDM model, $P_{MM}^{\CC}(a)$, we
define the parameter:
\begin{equation}\label{Fdef}
|F|\equiv\left| \frac{P_{MM}(a)
-P_{MM}^{\CC}(a)}{P_{MM}(a)}\right|_{a=1}=\left| \frac{\G^2(a)
-\G^2_{\CC}(a)}{\G^2(a)}\right|_{a=1}\,,
\end{equation}
where $\G(a)$ is the solution of (\ref{ggrowth4}) for the given DE
model with some effective EOS $\we=\we(a)$ (in some cases $\we$ can
be constant, but in general, it is a function of the scale factor),
and $\G_{\CC}(a)$ is the corresponding solution for $\we=-1$
($\CC$CDM). From its definition, the factor $F$ is a number,
computed for $a=1$ (i.e., the current Universe at $z=0$). On the
other hand, since we compare all models to the same observed galaxy
power spectrum $P_{GG}$, it follows that the above F-factor can be
directly related to the relative difference between the linear bias
factor of the given model with that of the $\CC$CDM model:
\begin{equation}
|F|=\left|1- \frac{P_{MM}^{\CC}(z)}{P_{MM}(z)}\right|_{z=0}=\left|1-
\frac{b^2(z)}{b^2_{\Lambda}(z)}\right|_{z=0} \label{Fbias}\,.
\end{equation}
As emphasized above, observations\,\cite{Cole05,Lahav02} show a
scale invariant linear evolution of $b^2_{\Lambda}(z)$ for the
$\Lambda$CDM model towards  $ b^2_{\CC}(0)=1.0\pm 0.1 $ at present.
Therefore, since all the DE models should, presumably, approach the
predictions of the successful $\Lambda$CDM model near the present
time, we require that any given DE model with a growth factor
$\G(a)$ should pass the following ``F-test''\,\cite{Ftest}:
\begin{equation}\label{Ftest}
|F|=\left|1-\frac{\G^2_{\CC}(a)}{\G^2(a)}\right|_{a=1}\leq 0.1\,.
\end{equation}
In performing the test,  both $\G(a)$ and $\G_{\CC}(a)$ must be
evolved from the recombination epoch, where the initial conditions
are fixed (see above), to the present time. It is understood that
the maximum limit, $F_{\rm max}=0.1$, is saturated when
$\G^2_{\CC}(1)/\G^2(1)=1.1$ and that the minimum limit, $F_{\rm
min}=-0.1$, when $\G^2_{\CC}(1)/\G^2(1)=0.9$.  In the next sections,
we apply these limits to the running $\CC$CDM and $\CC$XCDM models
in order to bound their respective parameter spaces.

\section{Effective method approach to the running $\La$CDM model\label{sect:runningLCDM}}

The F-test, defined in the previous section, is \textit{not} exactly
equivalent to requiring that $b^2(0)=1\pm 0.1$ for a given model,
but it is not very different from it and has the advantage of being
a relatively economical procedure. However, we need to check its
efficiency in some non-trivial situation before applying it to a
complex DE model, such as the $\CC$XCDM model. To this end, we first
apply the effective approach to the computation of matter
fluctuations, together with the F-test, to the running $\CC$CDM
model, for which we know the results of a complete
analysis\,\cite{FSS1}.

In the running $\CC$CDM model, the CC, or equivalently, its
associated energy density $\rL=\CC/8\pi\,G$, is an evolving
parameter. It ``runs'' because of the quantum loop effects of the
high energy fields and, therefore, it satisfies a renormalization
group (RG) equation. We refer the reader to the original literature
for details \,\,\cite{JHEPCC1,IRGA03,RGTypeIa}. Here, we just
highlight the basic concepts and equations, needed for our analysis.
In this framework, the physical RG running energy scale is
identified with $H$ (the Hubble parameter) and the solution of the
aforementioned RG equation reads
\beq \rho_\La(H,\nu)\,=\,\rho_\Lambda^0 +
\frac{3\,\nu}{8\pi}\,M_P^2\, \left(H^2(a;\nu)-H^2_0\right)\,,
\label{runlamb} \eeq
i.e., the DE density (in this case a running CC term) is an affine
quadratic law in $H$, where $\rho_\Lambda^0$ and $H_0$ are the
current values of these parameters. In this model we have a single
(dimensionless) new parameter $\nu$, given by
\begin{equation}
\nu\equiv \frac{\sigma}{12\,\pi}\,\frac{M^2}{M_P^2}\,, \label{nu}
\end{equation}
which is essentially the ratio (squared) of the effective mass $M$
of the high energy fields to the Planck mass ($M_P$), with fermions
contributing $\sigma=-1$ and bosons, $\sigma=+1$. If the effective
mass $M$ of the heavy particles, associated to some Grand Unified
Theory, is just the Planck mass $M_P$, the parameter $\nu$ takes the
value (positive or negative, depending on $\sigma$)
\begin{equation}
\nu_0=\frac{1}{12\,\pi}\simeq 2.6\times10^{-2}\,, \label{canonico}
\end{equation}
sometimes referred to as the ``canonical'' value.  In practice, the
preferred values of $\nu$ are smaller than $\nu_0$, as we will see,
which is, in fact, the natural situation because it corresponds to
having the heavy particles at some scale nearby, but below, the
Planck scale.
We point out that the Hubble function of the running model is also
$\nu$-dependent, $H=H(a;\nu)$. In the flat case,
\begin{eqnarray}\label{Hzzz}
H^2(a;\nu)= H_0^2\,\left[1+\Omega_M^0\,
\frac{a^{-3\,(1-\nu)}-1}{1-\nu}\right]\,.
\end{eqnarray}
For $\nu=0$, we recover the standard form corresponding to a
strictly constant $\CC$.

In the framework of the running $\CC$CDM model, there is energy
exchange between the vacuum and matter sectors and we have the
following mixed conservation law:
\begin{equation}\label{Bronstein}
{\rho}^{\prime}_{\CC}(a)+{\rho}^{\prime}_M(a)+\frac{3}{a}\,\rho_M=0\,.
\end{equation}
From (\ref{runlamb}) and (\ref{Bronstein}) we see that a
non-vanishing value of $\nu$ causes, not only a running of the CC
density  as a function of the scale factor (or the redshift),
$\rho_\La = \rho_\La(a;\nu)$, but also an exchange of energy between
$\rL$ and $\rM$. For $\nu=0$, however, $\rL$ becomes constant and
(\ref{Bronstein}) boils down to the old matter conservation law
(\ref{conslawmm}). These features are also apparent from
(\ref{runlamb}) and (\ref{Hzzz}).

For all the new dynamical features that a variable CC term may
entail, it should be clear that its EOS parameter still remains
$\wL=-1$. In this context, we may speak of the model as being
described within the ``CC picture''\,\cite{SS2}, that is to say, the
original formulation, in which the $\CC$ term is explicit and the
matter density is non-conserved. However, it may be advantageous to
perform a ``change of picture'', i.e., a description of the running
$\CC$CDM model within the ``DE picture''\,\cite{SS2}. In the latter,
we envision the given running CC model as if it were a DE model with
the same expansion history (that is, with the same numerical values
of $H$) but with self-conserved matter and DE densities
(\ref{conslawmm}) and (\ref{conslawde}). {The numerical matching of
the Hubble functions is essential in order to guarantee that the
physical results are the same in both pictures.  The basic reason
for moving into the DE picture is because we wish to find a
representation of our cosmological system where we can compute the
matter perturbation equations through Eq.\,(\ref{ggrowth3}), in
which the DE effects are confined only to the effective EOS
$\we=\we(a)$ and the ratio $r(a)=\rD(a)/\rM(a)$. Since, however,
(\ref{ggrowth3}) stems from the original Newtonian form
(\ref{ggrowth1}), which is derived under the hypothesis of total
matter conservation\,\cite{cosmobooks}, we must use a formulation of
our cosmological scenario in which this condition is also satisfied.
This representation is provided by the DE picture. }

While this alternate formulation is perfectly possible, the fact
that the matter density is non-conserved in the original CC picture,
suggests that the mapping of the latter into the DE picture can only
be carried out at the expense of finding a non-trivial effective EOS
parameter (actually some complicated function $\we=\we(a)$) relating
the self-conserved $\rD$ density and pressure $\pD$ in the DE
picture, $\pD=\we \rD$. This non-trivial EOS function for the
running $\CC$CDM model was determined in\,\cite{SS1,SS2}. For the
present purposes, it is convenient to first determine the normalized
DE density $\OD=\OD(z;\nu)$ explicitly and then apply (\ref{effw}).
{Following the procedure indicated above and adjusting the two
pictures such that the current values of $\OD$ and $\OL$ coincide,
we find}\,\footnote{For comparison, the corresponding normalized DE
density in the original CC picture is given by (\ref{runlamb}),
where $H$ has the explicit form (\ref{Hzzz}), divided by $\rco$. See
Refs.\,\cite{RGTypeIa} and \cite{SS2} for numerical plots of these
functions.}
\begin{eqnarray}\label{effDE}
\OD(a;\nu)=\OL^0+
\frac{\OM^0}{1-\nu}\,a^{-3}\left(a^{3\nu}-1+\nu\right)-
\frac{\nu\OM^0}{1-\nu}\,,
\end{eqnarray}
where $\OLo=1-\OMo$. (A generalization of this method allowing for
the mapping of any cosmological model with variable cosmological
parameters from the original ``CC picture'' into the ``DE picture''
was developed in \cite{SS2}, to which we refer the reader for
details on the entire procedure.) From this expression and
(\ref{effw}), we find the desired result,
\begin{equation}
\we
(a;\nu)=-1+(1-\nu)\,\frac{\left(a^{3\nu}-1\right)\OM^0\,a^{-3}\,}
{1-\nu-\OM^0+\left(a^{3\nu}-1+\nu\right)\OM^0\,a^{-3}\,}\,.
\label{eosltcdm}
\end{equation}
This effective EOS behaves nicely: $\we\to -1$ for $a\to 1$.
Expanding linearly the previous formula for small $\nu$ near our
epoch, we have
\begin{equation}\label{expwq}
\we (a;\nu)\simeq -1+3\,\nu\,\frac{\OM^0}{\OL^0}\,\,\frac{\ln
a}{a^3}\,.
\end{equation}
Since $a(t)<1$ for any look-back time, this equation shows that the
self-conserved density $\rD$ in the DE picture for this model does
not necessarily behave like quintessence ($\we\gtrsim -1$) in our
past; it is only so if $\nu<0$. It, however, behaves like ``phantom
DE''\,\cite{Phantom} (i.e. $\we\lesssim-1$) when $\nu>0$.

\begin{figure}
  \centering
\includegraphics[width=110mm]{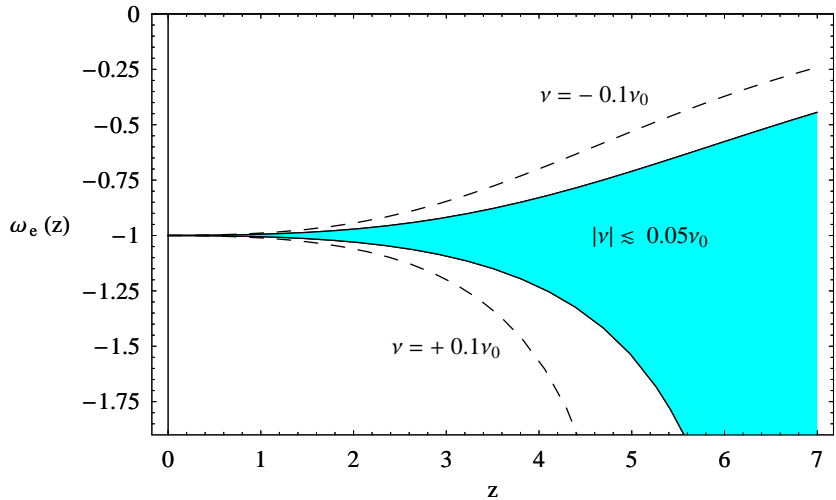}
\caption{The effective EOS for the running $\La$CDM model as a
function of the cosmological redshift $z$. We assume flat space
geometry and the prior $\Omega_M^0=0.3$. The shaded area satisfies
the ``F-test'' condition $|F|\leq 0.1$, see
Eq.\,(\protect\ref{Ftest}). }\label{weffigLtCDM}
\end{figure}
These features can be clearly identified in Fig.~\ref{weffigLtCDM},
where we show the evolution of the effective EOS for the running
$\La$CDM model, assuming a prior $\Omega_M^0=0.3$ for the normalized
matter density at present. For convenience, we display  $\we=\we(z)$
as a function of the cosmological redshift $z=(1-a)/a$. The
numerical examples shown in this figure correspond to four values of
$\nu$, expressed as small fractions of the canonical value
(\ref{canonico}). Specifically we plot (\ref{eosltcdm}) for $\nu=\pm
0.1\,\nu_0$ and $\nu=\pm 0.05\,\nu_0$. Although not shown in the
figure, from the analytic expression (\ref{eosltcdm}), it follows
that for $\nu>0$, the effective EOS for the running $\La$CDM model
goes to zero at very high redshift, whereas if $\nu<0$, $\we(z)\to
-\nu$ for $z\to\infty$. We note that if $\nu>0$, the denominator of
Eq.(\ref{eosltcdm}) vanishes at some redshift $z_1>0$. This is
related to the fact that for a positive $\nu$, the normalized DE
density (\ref{effDE}) vanishes at some point in the past. This is
easily seen from the sign of this function, which changes from
$\OD(a;\nu)<0$ in the far past ($\OD\to -\OMo/a^3<0$, for $a\ll1$)
to $\OD(a;\nu)>0$ at present ($\OD\to \ODo>0$, for $a\to 1$). This
transition is not possible for $\nu<0$. The divergence of $\we$ for
positive $\nu$ is not a real singularity of the theory since the
fundamental cosmological functions in both pictures, $\rM=\rM(a)$,
$\rL=\rL(a), \rD=\rD(a)$, $H=H(a)$, etc., are finite for all $a$.
The discontinuity of $\we=\we(a)$ is associated with the effective
description of the original model in the DE picture and has no
effect on our analysis. Indeed, the product function $\we(a)\,r(a)$
in the differential equations (\ref{ggrowth3}) and (\ref{ggrowth4})
remains finite for all $a$. Therefore, the computation of the growth
factor is free from singularities in the entire range of definition.

\begin{figure}
  \centering
\includegraphics[width=110mm]{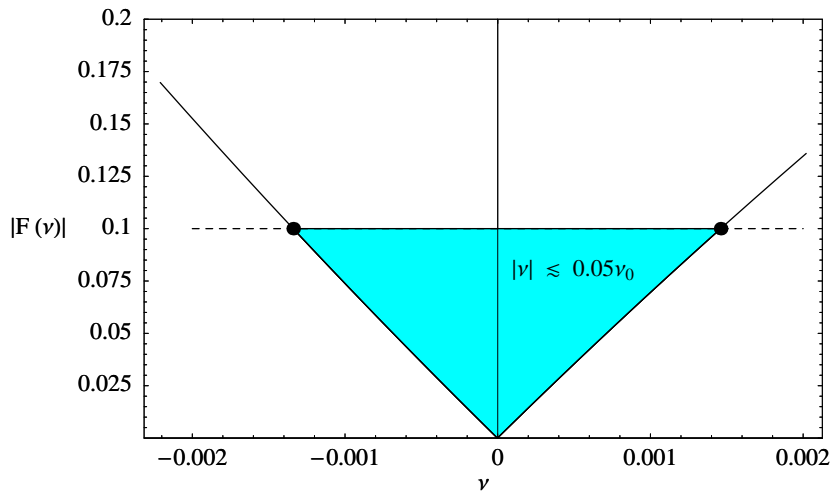}
\caption{F-test for the running $\CC$CDM model under the same
conditions as in Fig.\,\protect\ref{weffigLtCDM}. We show the factor
$F$, Eq.\,(\protect\ref{Fdef}), versus $\nu$. This parameter becomes
confined in the range $|\nu|\lesssim 0.05\,\nu_0\simeq 10^{-3}$, due
to the condition $|F|<0.1$, defined in (\protect\ref{Ftest}). The
horizontal dashed line corresponds to the maximum allowed deviation,
$|F|=0.1$, of the bias factor of this running model with respect to
that of the standard $\CC$CDM model.} \label{FfigLtCDM}
\end{figure}

Following the procedure explained in detail in section
\ref{sect:growth}, we have determined the growth factor for both the
standard $\CC$CDM and running $\CC$CDM models. For the former, we
naturally used $\we=-1$, and for the latter, Eq.\,(\ref{eosltcdm}).
After solving the perturbation equation (\ref{ggrowth4})
numerically, we computed the parameter F, defined in (\ref{Fdef}),
and applied the F-test (\ref{Ftest}), estimating the range of
allowed values for the parameter $\nu$, shown in
Fig.~\ref{FfigLtCDM}. The two values of $\nu$ which saturate the
F-test are the following: the upper bound, $F_{\rm max}=0.1$, which
corresponds to $\nu_{\rm max}\simeq - 1.3\times 10^{-3}$ and the
lower bound, $F_{\rm min}=-0.1$, which corresponds to $\nu_{\rm
min}\simeq + 1.4\times 10^{-3}$. They are represented by the two
black circles in Fig.~\ref{FfigLtCDM}. Therefore, the parameter
$\nu$ is confined to the range $|\nu|\lesssim 0.05\,\nu_0\cong
10^{-3}$, corresponding to the shaded areas in
Figs.~\ref{weffigLtCDM} and \ref{FfigLtCDM}. The other values,
$\nu=+0.1\nu_0$ and $\nu=-0.1\nu_0$, considered in
Fig.~\ref{weffigLtCDM}, correspond to $F_{\rm min}\simeq -0.17$ and
$F_{\rm max}\simeq 0.20$, respectively, and are excluded because
they are out of the allowed limits (the shaded region).

According to the 2dFGRS collaboration, the central values of the
normalized matter and DE densities at present, obtained from the
combined LSS and CMB data for the flat $\Lambda$CDM Universe, are
$\Omega_M^0=0.27$ and $\Omega_{\Lambda}^0=0.73$  respectively
\cite{Cole05}. These results are in good agreement with the most
recent analysis by the WMAP team, which reported $\Omega_M^0=0.28$
and $\Omega_{\Lambda}^0=0.72$ \cite{WMAP3Y}. To calculate the limits
on $\nu$ in this work, we considered flat space with the following
values of the matter and DE densities at present: $\Omega^0_M = 0.3$
and $\Omega^0_{D}=0.7$, respectively.

We end this section by noting that the limits we obtained for the
fundamental parameter of the running $\CC$CDM model,
\begin{equation}\label{limnu}
-10^{-3}\lesssim\nu\lesssim +10^{-3}\,,
\end{equation}
are essentially in agreement with previous estimates obtained by a
number of different methods and authors. In particular, the results
of \cite{vdecay} are in good agreement with (\ref{limnu}), although
in the latter study an indirect procedure was employed, which was
based on bounding the amplification of the matter density spectrum
in the recombination era caused by vacuum decay into CDM. We
emphasize that our results are also in good agreement with the full
calculation of coupled perturbations of matter, metric, and DE for
the running $\CC$CDM model, presented in \cite{FSS1}, where the DE
fluctuations were included as perturbations in the $\CC$ parameter.
A tighter bound for $\nu$ was obtained, $|\nu|\lesssim 10^{-4}$.
However, taking into account the economy of our present procedure,
as compared to \cite{FSS1}, we can assert that the effective
approach to the calculation of the matter density fluctuations, in
combination with the F-test, provides a reliable estimate of the
physical region of the parameter space.

From the tiny range of values of $\nu$ obtained, \,(\ref{limnu}), it
is patent that a procedure  based solely on the search for
non-standard features in the EOS profile (i.e. departures from the
featureless behavior $\we=-1$) can only be efficient if sufficiently
large redshift observations are used, which may be able to be
obtained from the DES and SNAP programs \cite{SNAP}. Even these
programs may have difficulty in differentiating the running
$\CC$XCDM model from the standard $\CC$CDM model, where $\we=-1.0$.
From Fig.~\ref{weffigLtCDM}, we see that the allowed interval for
the effective EOS parameter of this model at the redshift $z=1.7$
(the largest one reachable by SNAP) is rather small, and $\we$ does
not deviate appreciably from -1.0:
\begin{equation}\label{SNAPlimits}
-1.03\lesssim\we(z=1.7)\lesssim -0.97\,.
\end{equation}
The effective EOS parameter for the running $\CC$CDM model varies
very slowly with $z$. For example, at $z=2$,
$-1.05\lesssim\we(z=2)\lesssim -0.96 $. If a small departure from
the strictly constant $\we=-1$ could be observationally
substantiated, it would be a strong sign of dynamical behavior of
the DE. A remarkable feature of the running $\CC$CDM model is that
its effective EOS can mimic departures from the standard
cosmological model, both in the quintessence and phantom regimes,
even though the original model has nothing at all to do with
quintessence or with phantom-like fields. The original model (in the
``CC picture'') is nothing more or less than a model with a true,
albeit running, cosmological term and, hence,  with $\wL=-1$.

The most important result of our investigations of the running
$\CC$CDM model is that the LSS data are strongly sensitive to the
running of the $\CC$ term, which is, of course, the reason why the
parameter $\nu$ becomes strongly constrained. This confirms the
results of \cite{FSS1} within the present effective approach.
However, as noted earlier, in this model, the variability of $\CC$
is entangled with the non-conservation of matter. Therefore, in the
next section we investigate a model (the $\CC$XCDM model) with a
dynamical $\CC$ where matter is conserved, to see if the limits on
$\nu$ can be more relaxed. In the light of the successful
application of the effective method, we apply the same approach to
the $\CC$XCDM model.


\section{The framework of the $\La$XCDM model\label{sect:LXCDM}}

In the introduction, we have already explained the basic motivation
for the investigation of the $\Lambda$XCDM model, i.e., the
possibility of its being a solution to the cosmic coincidence
problem\footnote{We refer the reader to a detailed discussion of
this model in\,\cite{LXCDM1,GSSIRGAC12}.}. To achieve this important
property, the DE density in this model must have two components, to
wit: the running $\rho_{\Lambda}(a;\nu)$, which satisfies the RG
evolution equation (\ref{runlamb}), and the dynamical component $X$
(the ``cosmon''), whose underlying nature can be very general and
need not be specified here. These two components of the DE may be in
interaction. As already stated, $X$ is not necessarily related to
any scalar field. It could be some effective entity, related to the
structure of a more general theory, in which Einstein's gravity is
embedded, and which includes the effect of higher order terms. It is
supposed that the $X$ component behaves as a barotropic fluid with a
constant EOS parameter $\wX=p_X/\rho_X$. (In general, $\wX$ may not
be a constant, but a function of redshift.) Typically, the index
$\wX$ for the cosmon is in one of the following two expected ranges:
$\wX \gtrsim -1$ (quintessence-like cosmon) or $\wX \lesssim -1$
(phantom-like cosmon)\,\footnote{As noted in \cite{LXCDM1},
quintessence-like and phantom-like cosmons do not necessarily
exhibit the naively expected behaviors corresponding to quintessence
($d\rX/da<0$) and phantom energy ($d\rX/da>0$), respectively, since
in general there is an interaction between $X$ and the running
$\CC$, see (\ref{conslawDE2}). }. {Adopting the simplest possible
$\CC$XCDM scenario, we assume that the total matter density does
\emph{not} interact with the DE and that it is covariantly
conserved, thus satisfying (\ref{conslawmm}). If we define the total
DE density as the sum of the CC density and the cosmon density,
\begin{equation}\label{tDEX}
 \rho_D(a)=\rho_{\Lambda}(a)+\rho_X(a)\,,
\end{equation}
it follows from our assumption of matter conservation that $ \rho_D$
is also covariantly conserved. The cosmon X is actually defined
through this conservation condition\,\cite{LXCDM1}. Therefore, the
quantity (\ref{tDEX}) satisfies Eq.(\ref{conslawde}), which can be
recast as}
\begin{equation}
{\rho}^{\prime}_{\Lambda}(a)+{\rho}^{\prime}_X(a)\,+
\,\frac{3}{a}(1+w_X)\,{\rho}_X(a)=0\,. \label{conslawDE2}
\end{equation}
{From (\ref{conslawDE2}), it is clear that, although the total DE
density is locally conserved in the $\La$XCDM model, in general, the
individual $\rX$ and $\rL$ densities are not. There is a transfer of
energy between them, which is governed by the above equation.}

The expansion history of the $\CC$XCDM model is determined by its
Hubble function
\begin{equation}\label{HLXCDM}
H^2(a)=H_0^2\,\left[\OMo\,a^{-3}+\OD(a)\right]\,,
\end{equation}
where $\OD(a)=\rD(a)/\rco$ is the normalized total DE density
(\ref{tDEX}). The corresponding expression that satisfies the above
equations in the matter-dominated, flat Universe is given by
\begin{equation}
{\Omega_D}(a)=\frac{{\Omega_{\Lambda}^0}-\nu} {1-\nu}
+\frac{\epsilon\,\Omega^0_{M}\,a^{-3}}{w_X-\epsilon} +
\left[\frac{1-\OLo}{1-\nu}\,
-\frac{w_X\Omega^0_{M}}{w_X-\epsilon}\right]\,a^{-3(1+w_X-\epsilon)}\,,
\label{dewx}
\end{equation}
where, for convenience, we have defined
\begin{equation}
\epsilon\equiv\nu(1+w_X)\,. \label{epsilon}
\end{equation}
We will see, below, that this quantity must remain small for the
model to be compatible with primordial nucleosynthesis.

The normalized densities, at present, satisfy the relation
(\ref{friedeq}). For $a=1$ (i.e. $z=0$), it takes the form
\begin{equation}
\Omega_{D}^0+\Omega_M^0=\Omega_{X}^0+\Omega_{\Lambda}^0+\Omega_M^0=1\,,
\label{omflat}
\end{equation}
which is the current cosmic sum rule. With the help of this
relation, it is easy to see that, for $\nu=0$, the DE density
(\ref{dewx}) boils down to
\begin{eqnarray}
{\Omega_D}(a)=\Omega_{\Lambda}^0 + \Omega_{X}^0\,a^{-3(1+w_X)}\,.
\label{origquint}
\end{eqnarray}
Clearly, in this particular case, where $\nu$=0, the $\CC$XCDM model
mimics a system, consisting of a quintessence or phantom fluid
(depending on the value of $\wX$), together with a constant
cosmological term. From (\ref{origquint}), we then have three
possible scenarios:

\begin{itemize}
\item i) If $\wX\gtrsim -1$ and $\OXo>0$ (quintessence-like cosmon),
the expansion of the Universe can be stopped, provided that
$\OLo<0$, since the $X$ density gradually diminishes with the cosmic
time and the \textit{r.h.s.} of (\ref{origquint}) becomes negative.
Hence, there exists a future time, $a=a_{*}>1$, when $H(a_{*})=0$.
As mentioned previously, in the $\CC$XCDM model, the total
$\ODo\simeq 0.7$ (corresponding to $\OMo\simeq 0.3$). Therefore, due
to the contribution from $\OXo$ in the sum rule (\ref{omflat}), the
possibility that $\OLo<0$ should not be discarded a priori;

\item ii) If $\wX\lesssim-1$ and $\OXo>0$ (phantom-like cosmon), the
DE density increases indefinitely with the expansion. It does not
matter whether $\OLo$ is positive or negative, the Universe
unavoidably ends up in a super-accelerated phase, which triggers a
catastrophic disruption of all bound systems, a singularity known as
the ``Big Rip''\,\cite{Phantom};

\item iii) If $\wX\lesssim-1$ and $\OXo<0$,
the cosmon density, although phantom-like, acts with matter to
decelerate the expansion of the Universe. We have the opposite
situation to the Big Rip: the Universe becomes super-decelerated.
The kind of cosmon able to create this scenario was previously
called ``phantom matter''\,\cite{LXCDM1}. Phantom matter, therefore,
avoids the Big Rip and helps the Universe to reach $a=a_{*}>1$,
where $H(a_{*})=0$ (i.e., a stopping point). In the present
instance, this point will exist provided $\OLo>0$. Obviously phantom
matter is special in that it corresponds to negative energy density,
which is, however, not new in the literature\,\cite{McInness}. In
spite of its rather peculiar nature, phantom matter satisfies the
strong energy condition (see Fig.\,1 of \,\cite{LXCDM1}). As
previously discussed, the cosmon may well be an effective entity
and, therefore, could simulate the behavior of phantom matter. This
is in contrast to the ``standard phantom energy'', considered in the
previous case, which violates all of the classical energy conditions
and leads to a cosmic doomsday.

\end{itemize}

From the previous examples, with $\nu=0$, it is clear that there are
simple scenarios within the $\CC$XCDM model, in which the
cosmological expansion can be stopped at some point in the future.
The ``stopping'' point is actually a ``turning point'' in the
evolution of the Universe; it bounces back at that point and is
subsequently redirected towards the Big Crunch.  However, stopping
can be formulated on very general grounds within the parameter space
of the $\CC$XCDM model and is not restricted to $\nu=0$, as in the
previous examples. This issue is central to the cosmic coincidence
problem\,\cite{LXCDM1,LXCDM2} and is correlated with the existence
of a maximum of the ratio $r(a)$, defined in (\ref{ra}), which gives
the amount of DE energy versus matter at any time.  To further
address this problem, let us compute explicitly the function $r(a)$
in the $\CC$XCDM model in the matter dominated era. With the help of
(\ref{dewx}) and (\ref{rho}), we find
\begin{equation}
r(a)=\frac{(\OLo-\nu)\,a^{3}} {(1-\nu)\,\Omega^0_{M}}
+\frac{\epsilon\,}{(w_X-\epsilon)} +
\left[\frac{1-\OLo}{\OMo\,(1-\nu)}\,
-\frac{w_X}{w_X-\epsilon}\right]\,a^{-3(w_X-\epsilon)}\,.
\label{ratioLXCDM}
\end{equation}
In order to provide an acceptable explanation for the cosmic
coincidence problem, this ratio should be bounded and stay
relatively small throughout the entire history of the Universe. This
can be expressed through the condition
\begin{equation}
 \frac{r(a)}{r_0}<10\,,
 \label{ratiobound}
\end{equation}
where $r_0=\Omega_D^0/\Omega_M^0\simeq 7/3$ is the present value of
$r$ (of order one). The ratio (\ref{ratioLXCDM}) should, therefore,
reach a finite maximum in the evolution of the Universe. The
conditions for this to occur can be expressed as\,\cite{LXCDM1}
\begin{equation}
\frac{\OLo-\nu}{\wX\,(\OXo+\nu\,\OMo)- \epsilon\,(1-\OLo)}>0\,,\ \ \
\ \ \ \ (1 + w_X)\,\left(\Omega_{\La}^0-\nu\right)<0\,.
  \label{maximum}
\end{equation}
We can easily check that the simple stopping scenarios i) and iii),
mentioned above, are consistent with these requirements. We note
that $\wX=-1$ (i.e. a pure CC-like cosmon) is not allowed.

Besides the two conditions (\ref{ratiobound}) and (\ref{maximum}),
the ratio $r(a)$ should satisfy the nucleosynthesis constraint,
namely that its value at the primordial nucleosynthesis epoch should
be $|r_N|\lesssim 0.1$\,\cite{Ferreira97}. In that early epoch, the
ratio $r(a)$ is no longer given by (\ref{ratioLXCDM}) since we must
use the radiation equation for the (relativistic) matter density,
namely ${\Omega_R}(a)={\Omega^0_R}a^{-4}$ instead of (\ref{rho}).
Thus, we have:
\begin{eqnarray}\label{rzN}
r_N =
-\frac{\epsilon}{\wR-\wX+\epsilon}+\left[\frac{1-\OLo}
{\ORo(1-\nu)}-\frac{\wR-\wX}{\wR-\wX+\epsilon}\right]\,a_N^{-3\,(\wX-\epsilon)+1}\,,
\end{eqnarray}
where $\wR=1/3$ is the barotropic index for radiation and $a_N\sim
10^{-9}$, the scale factor at the nucleosynthesis time. Since
$\wX<-1/3$, the condition
\begin{equation}
|\epsilon| \lesssim 0.1\, \label{eb}
\end{equation}
insures that the contribution from the second term on the
\textit{r.h.s.} of (\ref{rzN}) is negligible and we find that
(\ref{eb}) is essentially equivalent to  $|r_N|\lesssim 0.1$.
Hereafter, we shall refer to (\ref{eb}) as the nucleosynthesis
bound.

For $\nu=0$ and $\OXo=0$ ($\OLo=1-\OMo$), Eq.\,(\ref{ratioLXCDM})
reduces to
\begin{equation}\label{rzLCDM}
r(a) =\frac{\OLo}{\,\OMo}\,a^{3}\,,
\end{equation}
which is, of course, the standard $\La$CDM model prediction for the
ratio $r(a)$. We see that, as the time passes, $a\to\infty$ and,
therefore, this ratio is unbounded from above, i.e., it can take any
arbitrarily large value. Thus, the cosmic coincidence problem boils
down to understanding why, at $a=1$ (the present time), the ratio is
just of order one. In other words, what makes our time special? This
question has no reasonable answer within the standard $\La$CDM
model. It also has no acceptable answer within the running $\La$CDM
model. Moreover, the ratio $r(a)$ in the running $\La$CDM model
\textit{cannot} be worked out as a particular case of the $\CC$XCDM
model because matter is not conserved in the former, whereas it is
conserved in the latter.

{Let us further elaborate on this ratio by considering its
evaluation within the running $\La$CDM for the two cosmological
pictures that we are considering, namely the CC and the DE pictures.
In the former, we have $r(a)=\rL(a)/\rM(a)$, where $\rM(a)$ and
$\rL(a)$ can be obtained from equations (7)-(9) of
Ref.\,\cite{IRGA03}, for example. The final result reads}
\begin{equation}\label{rarunLCDM}
r(a)_{\rm CC}
=\frac{\OLo}{\,\OMo}\,a^{3(1-\nu)}+\frac{\nu}{1-\nu}\,\left[1-a^{3(1-\nu)}\right]=
\frac{\OLo-\nu}{\OMo(1-\nu)}\,a^{3(1-\nu)}+\frac{\nu}{1-\nu}.
\end{equation}
{For $\OLo<\nu$, the CC density eventually becomes negative and, in
that case, there is stopping in the running $\CC$CDM. However, since
$\OLo\simeq 0.7$, this possibility entails a value of $\nu$, which
is ruled out by our result (\ref{limnu}). Hence, there is no viable
solution to the cosmic coincidence problem in this model. We wish to
stress that this result is independent of the particular
cosmological picture chosen to derive it. Indeed, in the DE picture
the ratio $r(a)$ is, instead, $r(a)=\rD(a)/\rM(a)=\OD(a)/\OM(a)$,
where $\OM(a)$ and $\OD(a)$ are given by (\ref{rho}) and
(\ref{effDE}) respectively. Therefore,}
\begin{equation}\label{rarunLCDMDE}
r(a)_{\rm DE}
=\left(\frac{\OLo}{\OMo}-\frac{\nu}{1-\nu}\right)\,a^3+\frac{a^{3\nu}-1+\nu}{1-\nu}
=\frac{\OLo-\nu}{\OMo(1-\nu)}\,a^3+\frac{a^{3\nu}}{1-\nu}-1\,,
\end{equation}
{where we used $\OMo=1-\OLo$. Notice that both ratios, $r(a)_{\rm
DE}$ and $r(a)_{\rm CC}$, satisfy the correct normalization at the
present time, i.e., $r(a=1)={\OLo}/{\OMo}$. Again, we see from
(\ref{rarunLCDMDE}) that the condition for the DE density to become
increasingly negative (leading to stopping) will occur only if
$\OLo<\nu$. Thus, we obtain the same conclusion as in the CC
picture.}

{The foregoing results indeed show that irrespective of the
cosmological picture used to perform the analysis, the conclusion is
the same, to wit: in the running $\CC$CDM model there is no natural
solution to the cosmic coincidence problem. Although this model does
provide some interesting dynamics for the CC term, it does not have
the ability to ameliorate the cosmic coincidence problem. It is only
when the X entity is introduced in interplay with a dynamical $\CC$,
that the ratio $r(a)$ takes the form of (\ref{ratioLXCDM}) and can
be kept within bounds throughout the entire history of the Universe,
which is indeed what is needed to solve the cosmic coincidence
problem.}

The three conditions,  (\ref{ratiobound}), (\ref{maximum}) and
({\ref{eb}}), define a limited 3D-region in the parameter space
$(\Omega_{\La}^0,\,w_{X},\,\nu)$ of the $\CC$XCDM model at a fixed
$\OMo\simeq 0.3$ \cite{LXCDM1}. In the next section, we give a
detailed numerical analysis of the $\CC$XCDM model, after including
two more conditions, both of which are related to the effective EOS
of the model. One of the conditions is its compatibility with the
LSS data. We express this compatibility condition, using the F-test
(\ref{Ftest}) on the linear bias parameter of the model, which
depends on the effective EOS approach to the calculation of the
growth factor. The other condition is the maximum allowed deviation
that we can tolerate for the value of the effective EOS parameter,
$\we=\we(z)$, away from the CC value, $\we= -1$, at $z=0$. As we
shall see, with these five conditions, we will be able to
significantly improve the determination of the physical volume in
the 3D parameter space of the $\CC$XCDM model, as compared to
\,\cite{LXCDM1}.


\section{Numerical analysis of the $\La$XCDM model: cosmic matter
perturbations versus cosmic coincidence\label{sect:numerical}}

In this section, we present a complete numerical analysis of the
$\La$XCDM model, in which its most salient feature is the inclusion
of the effective EOS approach to the growth of matter density
fluctuations, together with its ability to solve (or at least
significantly alleviate) the problem of cosmic coincidence,
mentioned in the introduction. The effective EOS for this model
is\,\cite{LXCDM1}
\begin{equation}\label{eEOS}
\we(a)=\frac{\pL+\pX}{\rL+\rX}=\frac{-\rL+\wX\,\rX}{\rL+\rX}=
-1+(1+\wX)\,\frac{\OX(a)}{\OD(a)}\,,
\end{equation}
where $\OD=\OD(a)$ is given by (\ref{dewx}), and
\begin{equation}\label{OXa}
\OX(a)=\left(\OXo-\frac{\nu}
{\wX-\epsilon}\,\OMo\right)\,a^{-3(1+\wX-\epsilon)} +\frac{\nu}
{\wX-\epsilon}\,\OMo\,a^{-3}\,.
\end{equation}
Note that for $\nu=0\,$, Eq.\,(\ref{OXa}) reduces to the expression
in the second term on the \textit{r.h.s.} of (\ref{origquint}), as
it should. Equations (\ref{OXa}) and (\ref{dewx}) may be used to
compute the effective EOS (\ref{eEOS}). Alternatively, (\ref{dewx})
may be substituted in (\ref{effw}), the two procedures being
equivalent since
\begin{equation}\label{check}
(1+\wX)\,\OX(a)=-\frac{a}{3}\,\frac{d\OD(a)}{da}\,.
\end{equation}
Clearly, the dynamical features of the cosmon play a preeminent role
in the behavior of the effective EOS for the $\CC$XCDM model. For
example, if the cosmon is quintessence-like ($\wX\gtrsim -1$ and
$\OX(a)>0$) near $a=1$, the overall effective EOS of the model will
be quintessence-like near our time. However, if the cosmon behaves
as phantom matter (viz.  $\wX\lesssim -1$ \textit{and} $\OX(a)<0$),
the effective EOS of the model will still be quintessence-like. On
the other hand, if the cosmon behaves as standard phantom DE
($\wX\lesssim -1$ and $\OX(a)>0$), the $\CC$XCDM model  will also
behave phantom-like as a whole. In general, the $\CC$XCDM model
behaves effectively as quintessence (phantom) near $z=0$, if and
only if $1+\wX$ \textit{and} $\OX$ have the same (opposite) signs.
This can be seen from \,(\ref{check}) and (\ref{eEOS}).

\begin{figure}[t]
  \centering
\includegraphics[width=70mm]{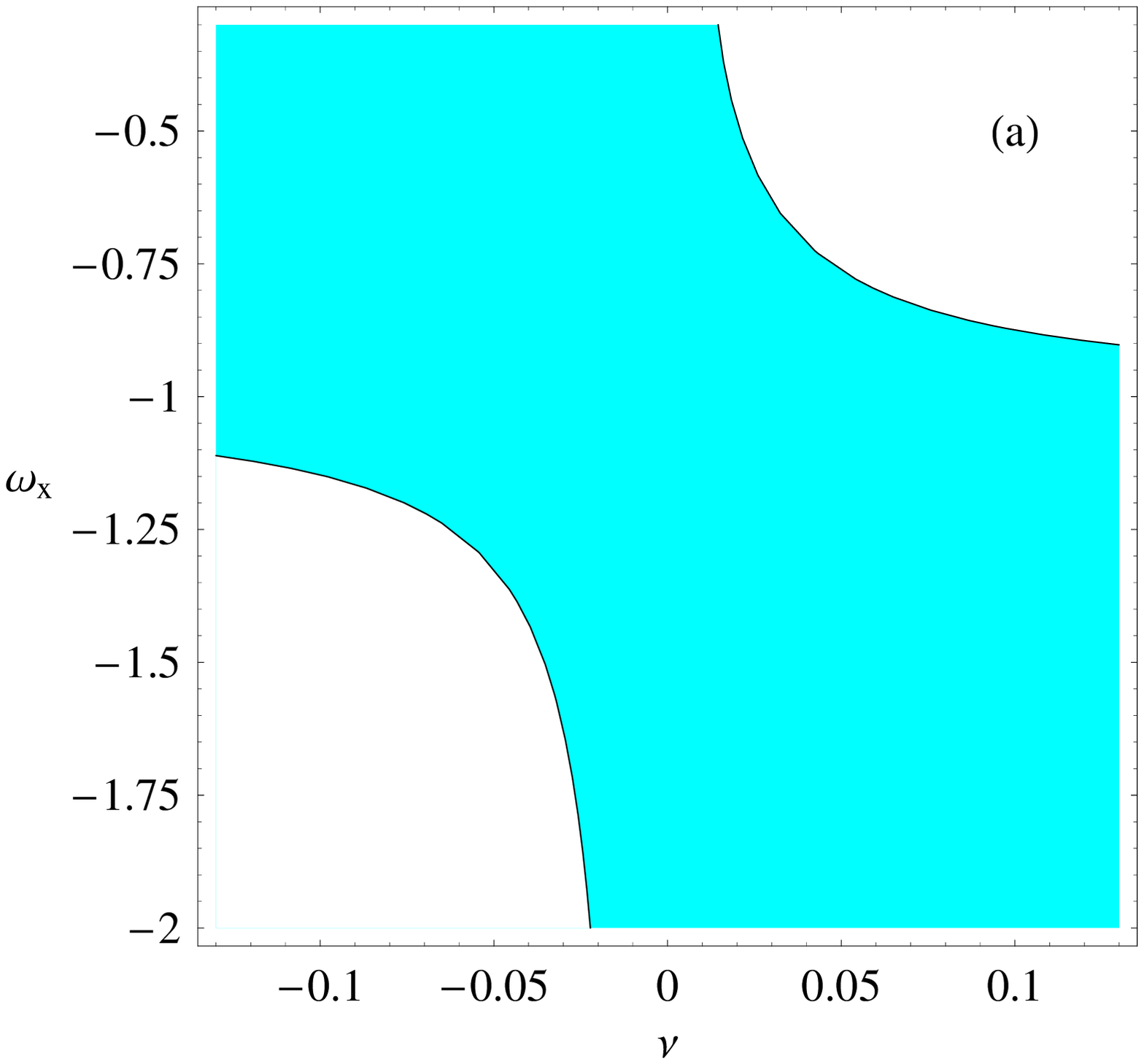}\hspace{2mm}
\includegraphics[width=70mm]{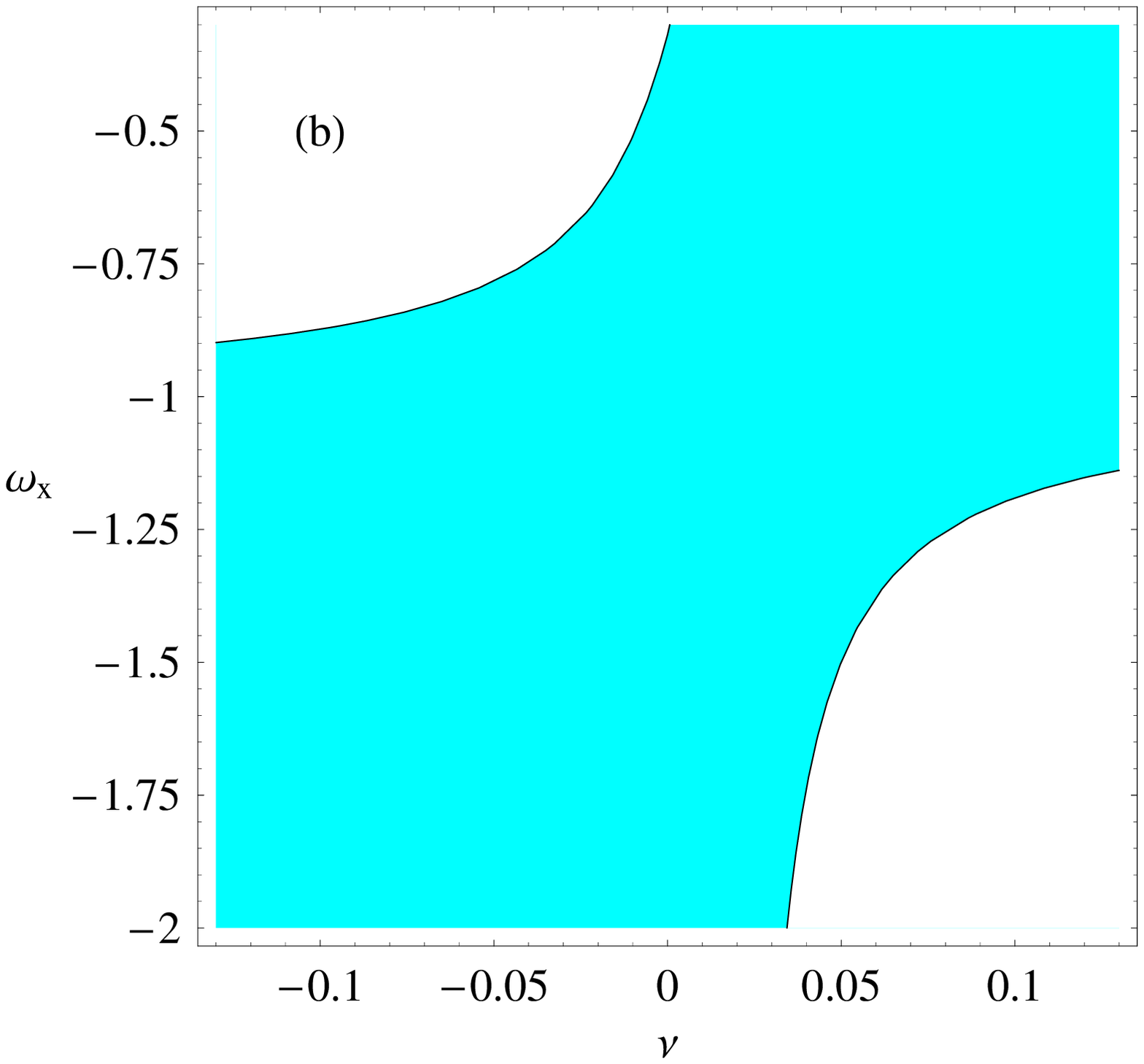}\hspace{2mm}
\includegraphics[width=70mm]{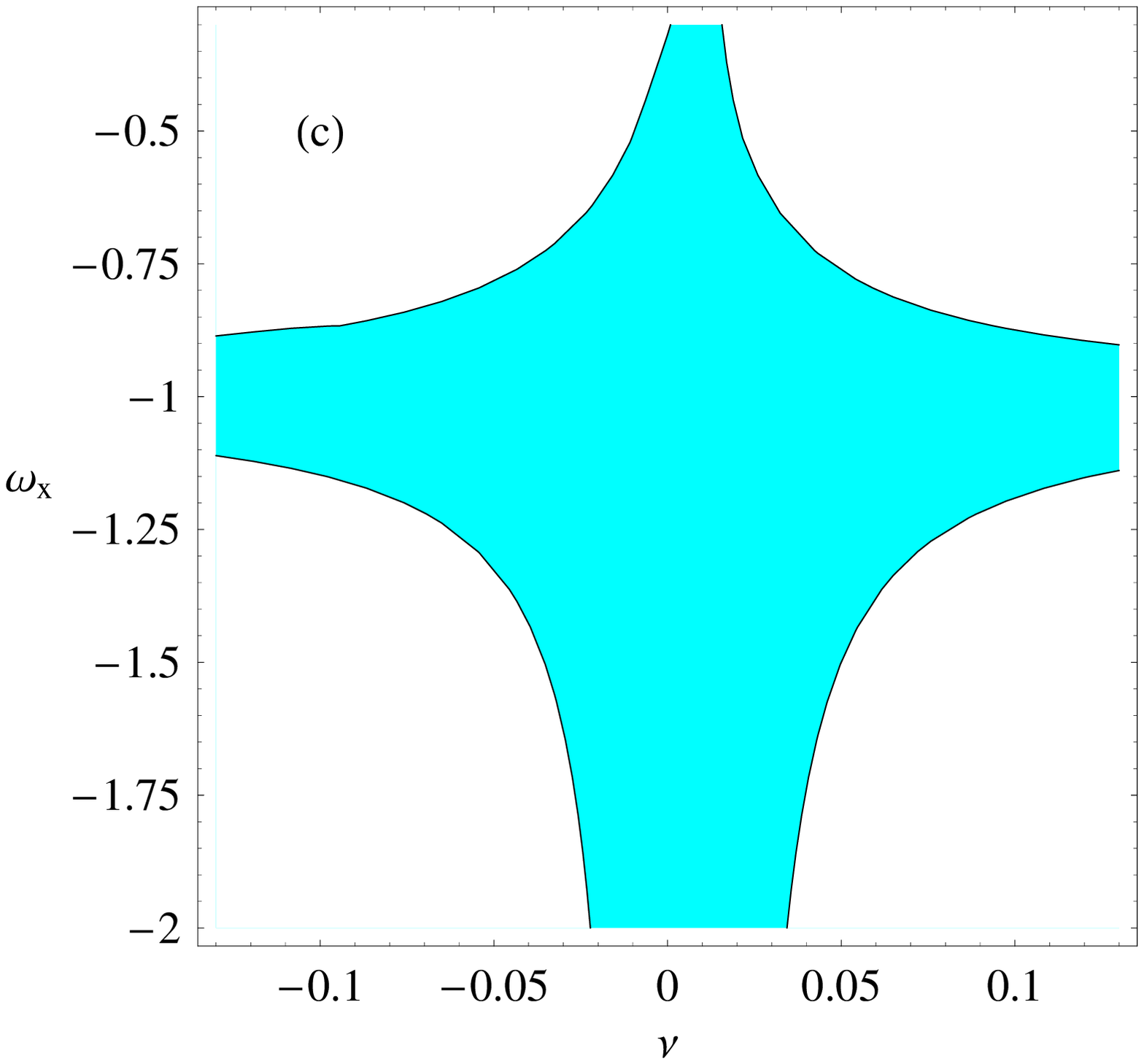}
\caption{F-test condition (\protect\ref{Ftest}) for the $\La$XCDM
model assuming flat geometry with $\OMo=0.3$ and individual DE
densities $\OLo=0.65$ and $\OXo=0.05$. Shown are: a) region of the
$(\nu,w_X)$ plane in which $F<0.1$; b) the corresponding region
where $F>-0.1$; and c) the allowed region by the F-test ($|F|<0.1$),
i.e. the intersection of regions a) and b) . } \label{fcondLXCDM}
\end{figure}

For the numerical analysis, we insert the EOS formula (\ref{eEOS})
into (\ref{ggrowth4}) for the effective growth of the density
perturbations. We then solve this equation and perform the F-test
(\ref{Ftest}). That is, we impose a condition on structure
formation, whereby we discard all points of the parameter space, for
which the growth factor of our model at $z=0$ deviates by  more than
$10\%$ from that of the $\Lambda$CDM.  In the analysis, we have to
also include the three conditions discussed at the end of the last
section, namely, the primordial nucleosynthesis constraint, the
stopping condition, and the bounding condition on the ratio
$r=r(a)$. We begin the analysis by checking the ability of the
F-test, alone, (\ref{Ftest}), to define a limited region of the
$(\wX,\nu)$ plane. As mentioned previously, we assume a flat
Universe with $\OMo=0.3$ and $\ODo=0.7$. For the sake of
illustration, we split the DE density into its individual components
as follows: $\OLo=0.65$ and $\OXo=0.05$ (which, of course, add up to
the value of $\ODo$). In Fig.\,\ref{fcondLXCDM}, we show the regions
of ($w_X, \nu$) plane that fulfill the partial constraints: $F<0.1$
(Fig.~\ref{fcondLXCDM}a) and $F>-0.1$ (Fig.~\ref{fcondLXCDM}b), as
well as the more restrictive region $|F|<0.1$
(Fig.~\ref{fcondLXCDM}c), which represents their intersection. All
points in these figures automatically satisfy the nucleosynthesis
bound (\ref{eb}). This means that, in this case, the F-test is
already more restrictive than the nucleosynthesis bound.

We  note that $w_X$ is less constrained when $\nu=0$, i.e., when the
$\CC$XCDM model behaves as a quintessence (or phantom) model with a
cosmological constant --  see \,(\ref{origquint}). In this case, we
have $-14\lesssim w_X \lesssim  -0.3$. Similarly, when $\wX$ is very
close to the value $\wX=-1$, the cosmon behaves as a cosmological
constant and the range of $\nu$ is almost unconstrained by the
F-test. This scenario effectively corresponds to having a strict
cosmological constant (in this case, the cosmon), together with a
variable cosmological term, $\CC$. These unconstrained situations
will change dramatically when the other restrictions (particularly
that of \textit{non}-cosmic coincidence) are also imposed in
combination with the F-test. We have already seen, in the previous
section, that $\wX=-1$ is actually forbidden. However, the F-test,
alone, (as a strategy to determine the principal restrictions due to
structure formation) is able to highly constrain regions, which the
other conditions are not able to do, as we will see below.

\begin{figure}[t]
  \centering
\includegraphics[width=90mm]{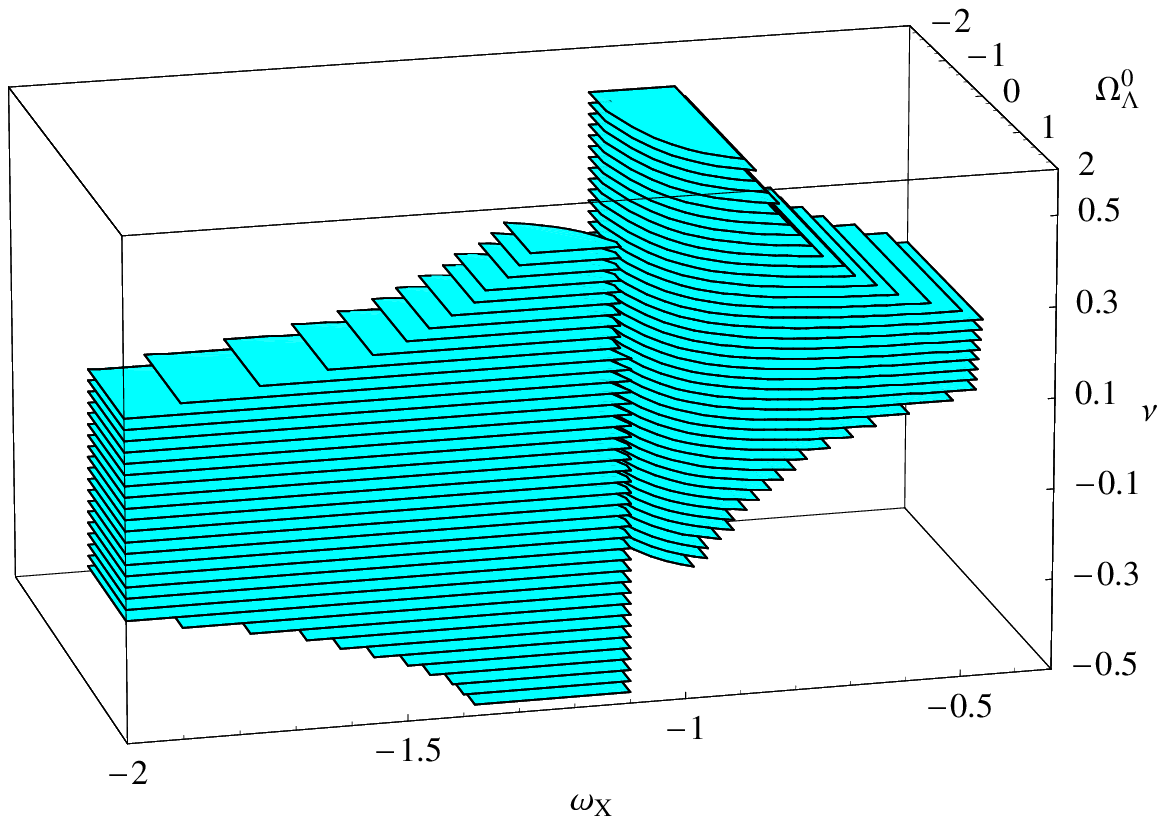}\hspace{2mm}
\includegraphics[width=88mm]{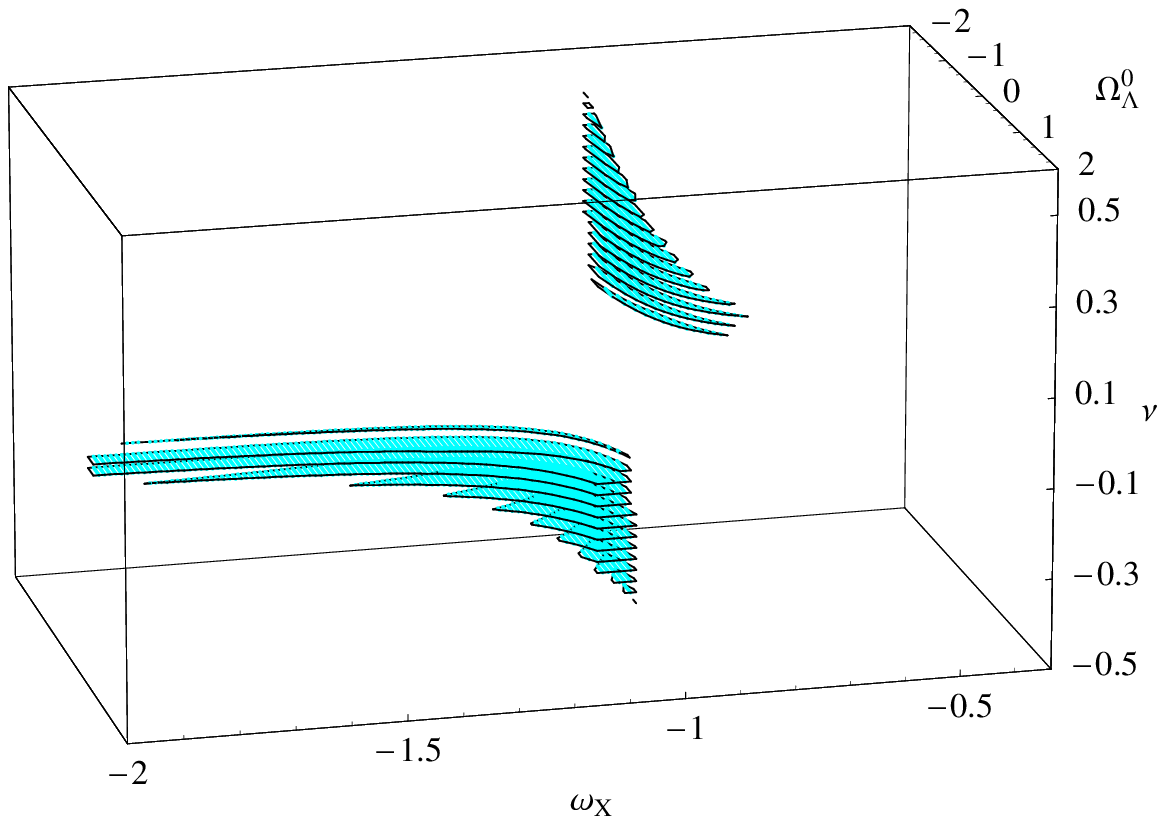}
\caption{Upper panel: Volume of points allowed by \textit{Conditions
1-3} described in the text, i.e. nucleosynthesis plus the two
restrictions associated with the solution of the coincidence
problem; Lower panel: Volume of points allowed by \textit{all}
constraints discussed in the text, namely \textit{Conditions 1-5},
which include the LSS constraint and the EOS restriction at $z=0$.
We assume flat space geometry with $\OMo=0.3$ and $\ODo=0.7$.}
\label{fig5}
\end{figure}

By assuming that $\Omega_M^0=0.3$ for the present matter content of
the Universe (which can be deduced from LSS observations alone) and
that the Universe is flat, the $\Lambda$XCDM model is left with
three free parameters, namely, $\wX$, $\nu$ and $\OLo$. The cosmon
density $\OXo$ is, then, no longer independent, since it is fixed by
the cosmic sum rule (\ref{omflat}). In a previous
work\,\cite{LXCDM1}, the parameter space of the $\CC$XCDM model was
constrained by imposing the three conditions (\ref{ratiobound}),
(\ref{maximum}), and ({\ref{eb}}), emerging from three important
physical requirements:

\begin{itemize}
\item {\it Condition 1: Nucleosynthesis bound.}

The expansion rate depends directly on the amount of DE. If we are
not to spoil the standard Big Bang predictions (about, e.g., light
element abundances), the DE density at the nucleosynthesis time
should not be very large. Specifically, it is required
that\,\cite{Ferreira97}
\begin{equation}\label{randrD}
\tilde{\Omega}_D(a_N)=\frac{r_N}{1+r_N}\lesssim 0.1\,,
\end{equation}
which is essentially equivalent to
\begin{equation}
|r_N|\equiv\left|\frac{\rho_D(z)}{\rho_M(z)}\right|_{(z=z_N)}\lesssim
0.1\,,
\end{equation}
where $z_N\sim10^9$ is the cosmological redshift at the
nucleosynthesis era.

\item {\it Condition 2: Stopping condition.}

As commented in section \ref{sect:LXCDM}, in the $\Lambda$XCDM
model, the ratio $r(a)$ of the DE to matter density may exhibit a
maximum. This feature is related to a future stopping and subsequent
reversal of the expansion of the Universe, expressed by
(\ref{maximum}).

\item {\it Condition 3: Low maximum of the ratio $r(a)$.}

For a solution, or at least  a substantial alleviation, of the
coincidence problem, we must further require that the stopping point
of the expansion of the Universe is preceded by a sufficiently small
maximum value of the ratio $r(a)$, defined in (\ref{ra}). In the
standard $\Lambda$CDM model, $r(a)$ is unbounded and its present
value, $r_0\sim1$, is related to the recent transition from
decelerated to accelerated expansion. Thus, within the standard
$\Lambda$CDM, we can conclude that we are living in a very special
epoch, i.e., very close to the transition epoch. Alternatively, if
$r(a)$ remains bounded and sufficiently small for the entire history
of the Universe, as can be the case in the $\Lambda$XCDM model, the
fact that $r_0\sim 1$ should no longer be regarded as a coincidence
since the relation $r\sim 1$ could hold for most of the lifetime of
the Universe. Therefore, in order to accommodate this appealing
feature in the $\CC$XCDM model, we search for points in the
parameter space that not only allow for the existence of a maximum
for $r(a)$, but also those points, for which the value of the
maximum is sufficiently small. Specifically, we express
quantitatively this condition by \,(\ref{ratiobound}).

Conditions 2 and 3 are actually related since the existence of a
maximum for $r(a)$ is correlated with the existence of stopping. We
have, nevertheless, kept them separated in order to stress that the
stopping condition is not sufficient to solve the cosmic coincidence
problem, it is only a necessary condition. We still have to demand a
``qualified stopping'', i.e., one preceded by a sufficiently low
maximum. Conditions 2 and 3 could then be unified and be
collectively referred to as \textit{The ability to solve the cosmic
coincidence problem}.

\end{itemize}
The subset of points satisfying these three conditions was found in
\cite{LXCDM1} to constitute a volume in the 3D parameter space,
shown in Fig.~\ref{fig5} (upper panel). In this paper, we have
incorporated a fourth and very important condition in the numerical
analysis:

\begin{itemize}
\item {\it Condition 4: Consistency with the data on structure formation.}

We have already explained in detail, throughout this work, the way
in which we have included this condition in our analysis, namely,
through the effective EOS approach and the implementation of the
F-test,\,(\ref{Ftest}). This has been one of the principal aims of
this work.

\end{itemize}

There is yet one more observational requirement that can be
demanded:

\begin{itemize}
\item {\it Condition 5: EOS condition at $z=0$}

The value of the effective EOS parameter should behave as a CC in
the recent past. However, the effective EOS of the $\CC$XCDM model
(\ref{eEOS}) does \textit{not} automatically satisfy the  ``CC
condition'', $\we(z)\to -1$ for $z\to 0$, as does the effective EOS
of the running $\CC$CDM model (\ref{eosltcdm}). Therefore, we wish
to make sure that this condition is indeed satisfied by the
$\CC$XCDM model, within the limits of the latest observational data.
We normalize this EOS bound at $z=0$, which is elaborated below. The
quantitative restriction associated with this condition is
\begin{equation}
|1+\omega_e(z=0)|\leq 0.3\,. \label{EOSCond}
\end{equation}

\end{itemize}

This can be justified from recent experiments, which strongly
suggest that the EOS parameter should be close to $-1$. For
instance, the combination of WMAP and the Supernova Legacy Survey
(SNLS)\,\cite{SNLS} data (under the assumption of spatial flatness)
yields\,\cite{WMAP3Y}:
\begin{equation}
\omega_e=-0.967^{+0.073}_{-0.072}\,.
\end{equation}

Even without the prior that the Universe is flat, the value of the
EOS parameter preferred by WMAP, large-scale structure and
supernovae data is still very close to that of a cosmological
constant\,\cite{WMAP3Y},
\begin{equation}
\omega_e=-1.08\pm 0.12\,.
\end{equation}
These observational results assume that the EOS parameter is
constant and, therefore, they are not directly applicable to the
$\Lambda$XCDM model. Nevertheless, we prefer to adopt a conservative
point of view and impose the additional constraint from
(\ref{EOSCond}), which we referred to as condition 5, to our model.

\begin{figure}[h!]
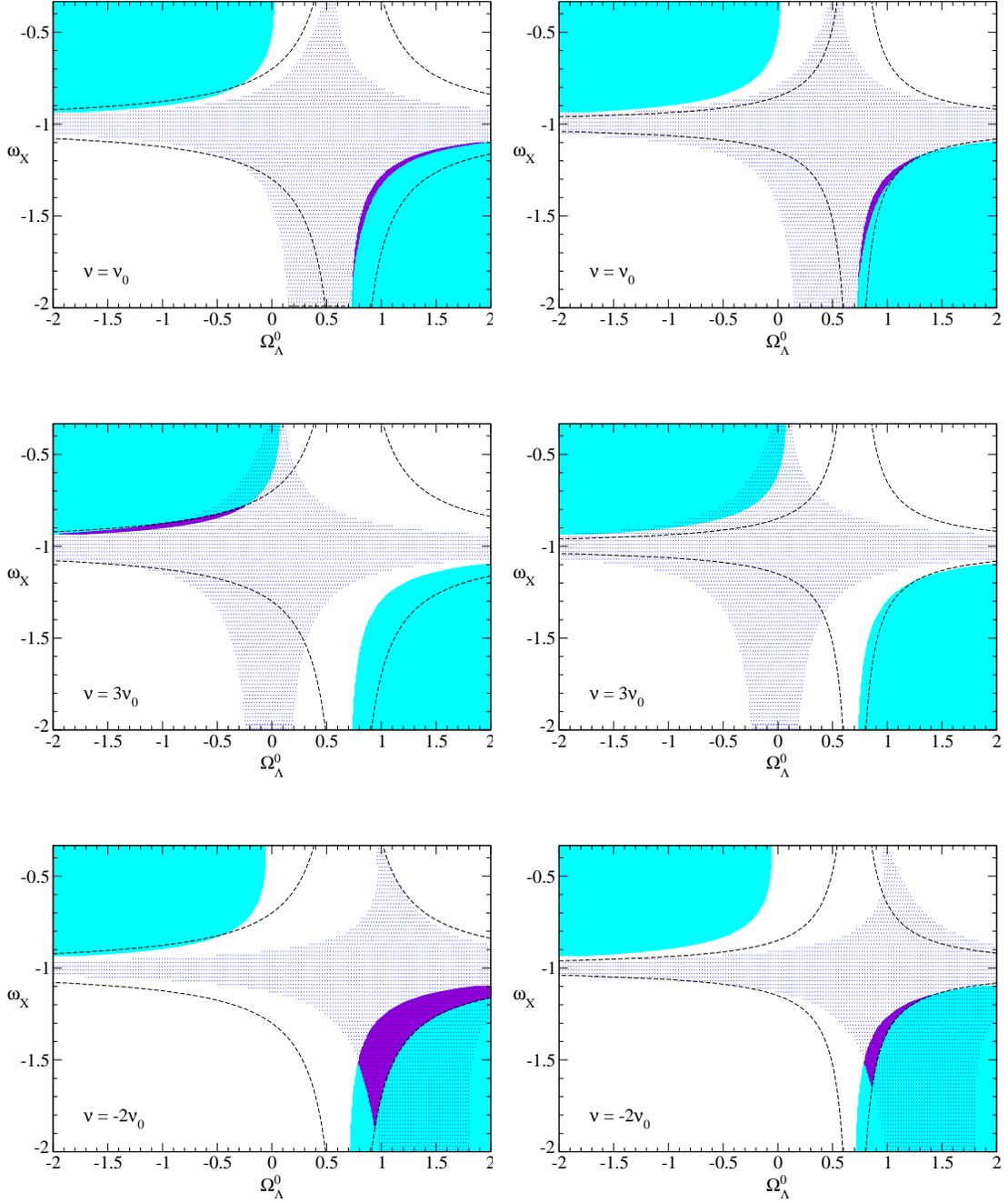

  \centering
\includegraphics[width=72mm]{Figures/figure51a.eps}\hspace{2mm}\includegraphics[width=72mm]{Figures/figure51b.eps}\\[5ex]
\includegraphics[width=72mm]{Figures/figure52a.eps}\hspace{2mm}\includegraphics[width=72mm]{Figures/figure52b.eps}\\[5ex]
\includegraphics[width=72mm]{Figures/figure53a.eps}\hspace{2mm}\includegraphics[width=72mm]{Figures/figure53b.eps}
\caption[3D plots]{$\nu$-slices of the physical 3D region of the
$\Lambda$XCDM parameter space (Fig.\,\protect\ref{fig5}, lower
panel) for $\nu=\nu_0,\ 3\nu_0$ and $-2\,\nu_0$.
 Points in the solid-shaded area fulfil the nucleosynthesis bound
 and the two conditions associated with the solution of the coincidence
problem (\textit{Conditions 1,2 and 3} in the text).
 The dotted region consists of
 points allowed by the LSS data (\textit{Condition 4}).
 Points inside the dashed lines satisfy \textit{Condition 5}: $|1+\omega_e(0)|\leq 0.3$ (left panel)
 and $|1+\omega_e(0)|\leq0.15$ (right panel).
 The physical region (the darker one) in each case is the common overlap.
 Note that quintessence-like cosmons are
forbidden if $|1+\omega_e(0)|\leq 0.15$.}\label{fig6}
\end{figure}

Remarkably enough, after imposing the five conditions listed above,
we are still left with a non-empty volume of points in 3D parameter
space that satisfy all of them. This can be seen in
Fig.~\ref{fig5}\,(lower panel). Comparing with the upper panel of
this figure, we see that the new constraints greatly diminish the
final volume of points allowed in the parameter space. This means
that conditions 4 and 5 are very restrictive when combined with the
first three conditions.

\begin{figure}[t]
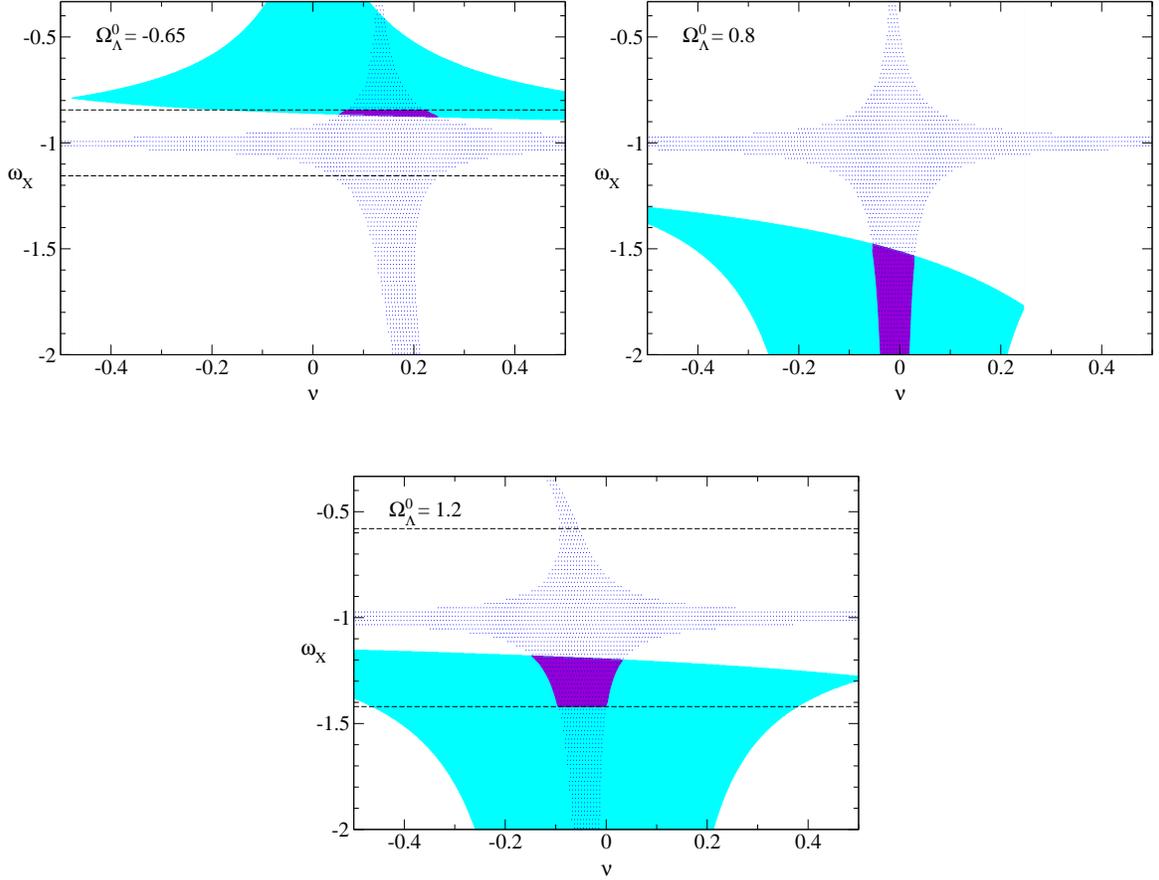

\centering
\includegraphics[width=75mm]{Figures/figure6a.eps}\hspace{3mm}\includegraphics[width=75mm]{Figures/figure6b.eps}\\[5ex]
\includegraphics[width=75mm]{Figures/figure6c.eps}
\caption{Three $\OLo$-slices of the physical 3D region of the
$\Lambda$XCDM parameter space for $\OLo=-0.65,\,0.8,\,1.2$. The
shaded region is allowed by \textit{Conditions 1-3} in the text. The
dotted one is allowed by \textit{Condition 4} (the F-test), and the
region between the vertical dashed lines is allowed by
\textit{Condition 5}. The $\OLo=0.8$ plot is not restricted by this
condition. The darker areas in the three plots are the final allowed
regions.}\label{fig7}
\end{figure}

\begin{figure}[t]
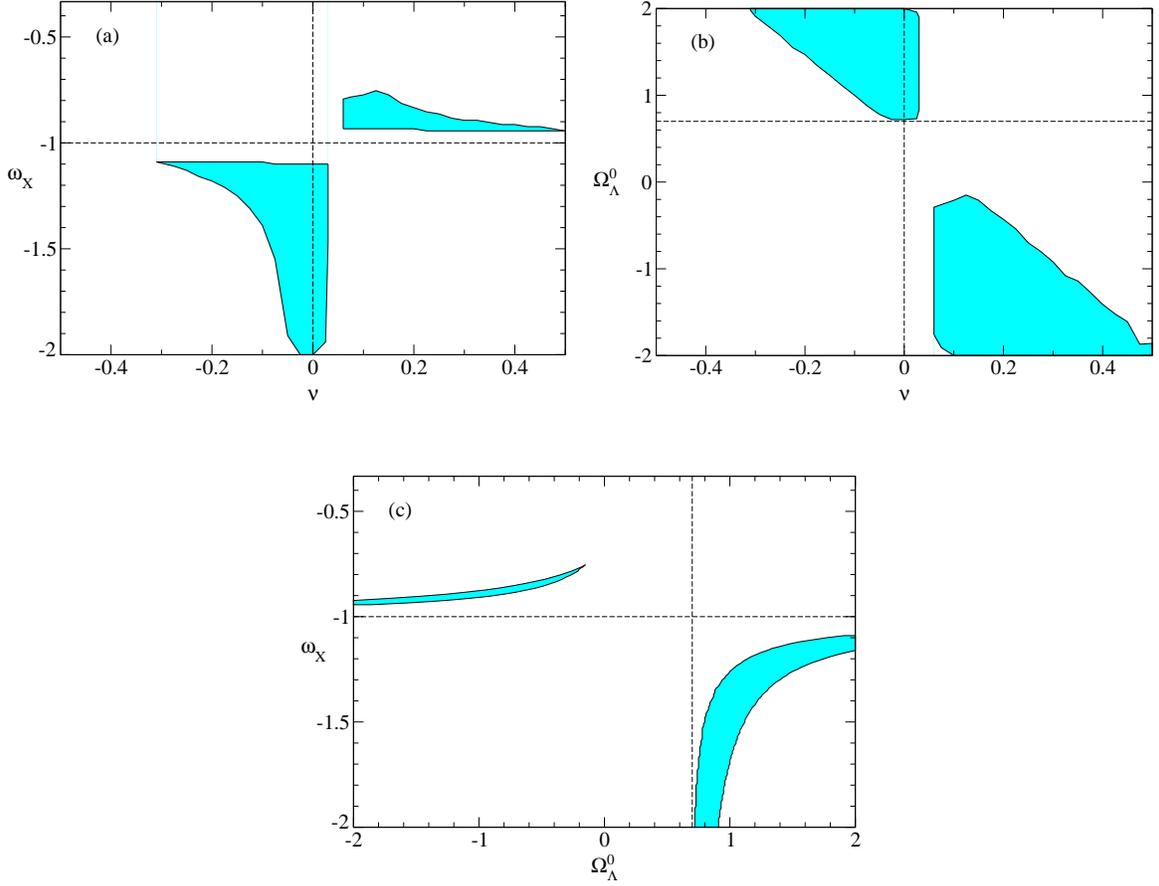

\centering
\includegraphics[width=75mm]{Figures/figure7a.eps}\hspace{3mm}\includegraphics[width=75mm]{Figures/figure7b.eps}\\[5ex]
\includegraphics[width=75mm]{Figures/figure7c.eps}
\caption{Projections of the final 3D physical region of the
$\CC$XCDM model (Fig.\,\protect\ref{fig5}, lower panel) onto the
three different planes. The vertical and horizontal dashed lines
mark the values $\nu=0$, $\wX=-1$ and $\OLo=+0.7$ wherever any of
these parameters appear.}\label{fig8}
\end{figure}

If we wish to better visualize the effect of the different
constraints, it is more convenient to study two-dimensional slices
of the final 3D volume in Fig.~\ref{fig5} (lower panel). In order to
do this, we fix one of the three free parameters. In
Fig.~\ref{fig6}, we plot three different $\nu$-slices of the final
volume: $\nu=\nu_0,\, 3\nu_0,\, -2\nu_0$, where $\nu_0$ is the
canonical value, defined in (\ref{canonico}). In Fig.~\ref{fig7}, we
plot three $\OLo$-slices: $\OLo= -0.65, \,0.8, \,1.2$. Let us
discuss the $\nu$-slices and restrict ourselves, for the moment, to
the panel on the left of Fig.~\ref{fig6}. We see that both signs of
$\nu$ are possible and that the cosmon may be both quintessence-like
($\wX\gtrsim -1$) and phantom-like ($\wX\lesssim -1$), although the
former is only possible for positive $\nu$, according to
Fig.~\ref{fig6} (middle figure on the left).  This feature does not
depend on this particular $\nu$-slice ($\nu=3\nu_0$), it is general.
It can be confirmed from Fig.~\ref{fig8}, where the projections of
the bulk volume of final points (lower panel of Fig.~\ref{fig5})
onto the three possible planes are shown. Also general is the
property that phantom-like cosmons are compatible with positive
(greater than +0.7) values of $\OLo$, whereas quintessence-like
cosmons demand a negative energy density of the vacuum (cf.
Fig.~\ref{fig8}).

Since the precision of observational data will increase in the
future, we have also studied what happens if we make the condition
on $\omega_e(0)$ more stringent than condition (\ref{EOSCond}),
namely,
\begin{equation}\label{tighterEOS}
|1+\omega_e(0)|\leq0.15\,.
\end{equation}
The result of applying this tighter constraint is that the volume of
points allowed becomes further reduced, as can be seen in the panel
on the right of Fig.~\ref{fig6}. The most conspicuous effect is
that, now, the $X$ component can only be phantom-like ($\wX\lesssim
-1$); quintessence-like cosmons ($\wX\gtrsim -1$) are no longer
permitted.

Consider now the $\OLo$-slices ($\OLo= -0.65, \,0.8, \,1.2$) of the
final volume in Fig.~\ref{fig7}. In general, only a small subset of
all the points allowed by the F-test is allowed by the full set of
constraints. Conversely, the area allowed by the old constraints
(\textit{Conditions} 1, 2 and 3) is highly restricted by the F-test.
For example, the $\OLo= -0.65$ slice shows a critically constrained
case, where the concurrence of the five conditions, particularly the
EOS restriction at $z=0$ (\textit{Condition 5}), leave a very small
area allowed in that slice.

Let us now consider the projections of the final allowed volume in
Fig.~\ref{fig8}. The signs of $\nu$ and $\OLo$ are correlated, as
shown in Fig.~\ref{fig8}b. Negative values of $\OLo$ are compatible
only with positive values of $\nu$. Similarly, positive values of
$\OLo$ admit mostly negative values of $\nu$, although there is
still a range of $\nu>0$ allowed values, which are of order of the
canonical value $\nu_0\sim 10^{-2}$. From Fig.~\ref{fig8}c, we see
that the signs of $1+\wX$ and $\OLo$ are also correlated. Thus,
quintessence-like cosmons ($1+\wX\gtrsim 0$) demand that $\OLo<0$,
whereas phantom-like ones ($1+\wX\lesssim 0$) require that
$\OLo>+0.7$. However, it should be stressed that if we would apply
the tighter condition, indicated in \, (\ref{tighterEOS}), only the
phantom-like character is possible for the cosmon; in other words,
the thin quintessence regions in Fig.~\ref{fig8}a and
Fig.~\ref{fig8}c then disappear altogether.

The reduction of the original region, permitted by the first three
conditions, to the final region, filtered by all the five
conditions, is best seen, comparing the projection plots in
Fig.\,~\ref{fig8} with the old projection plots in Fig.\,3 of
\,\cite{LXCDM1}. For instance, the present range of values for the
$\nu$ parameter become bounded from above and below and it is no
longer possible to have $|\nu|>0.5$. Another dramatic restriction
that occurs in the present analysis is the near exclusion of the
region where the cosmon is quintessence-like, the very small region
of the allowed parameter space where $\wX\gtrsim -1$. In this
region, $\nu$ cannot be negative or zero (or even close to zero) and
must be $\nu\gtrsim 0.1$.  This is clearly seen in the projection
plot of Fig.\,~\ref{fig8}a. We conclude that in most of the
parameter space, the cosmon $X$ behaves phantom-like ($\wX\lesssim
-1$). Since, however,  $\OLo>+0.7$ in this region and the current
value of the total DE density is fixed at $\ODo=+0.7$, the cosmic
sum rule (\ref{omflat}) indicates that in most of the allowed
parameter space, $\OXo<0$. Put another way, the cosmon behaves
mostly as phantom matter. This was to be expected since phantom
matter satisfies, as noted previously, the strong energy
condition\,\cite{LXCDM1}, and therefore it effectively behaves like
additional matter, helping to stop the expansion of the Universe at
some time in the future.

\begin{figure}[t]
 \centering
\includegraphics[width=100mm]{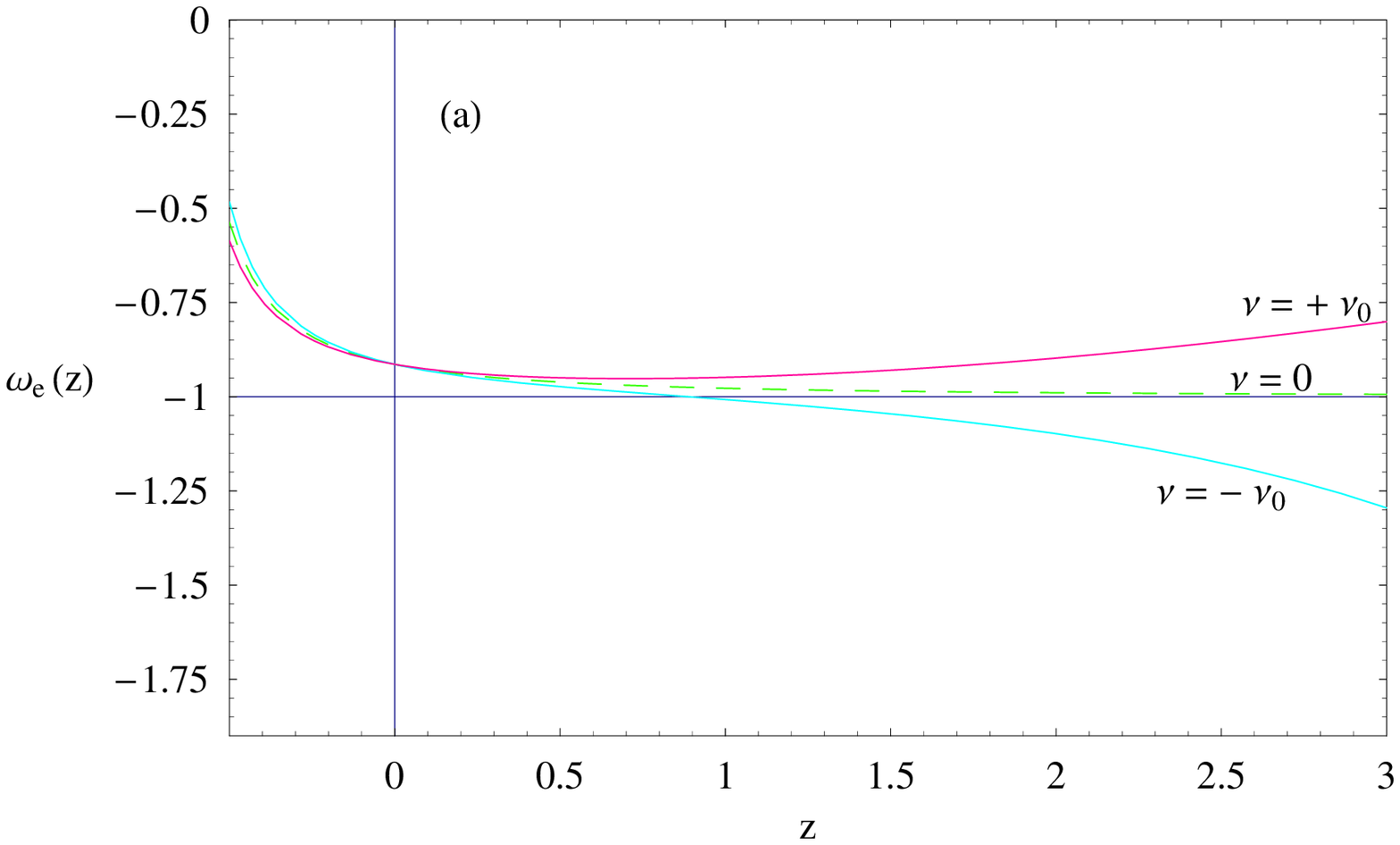}\\
\includegraphics[width=100mm]{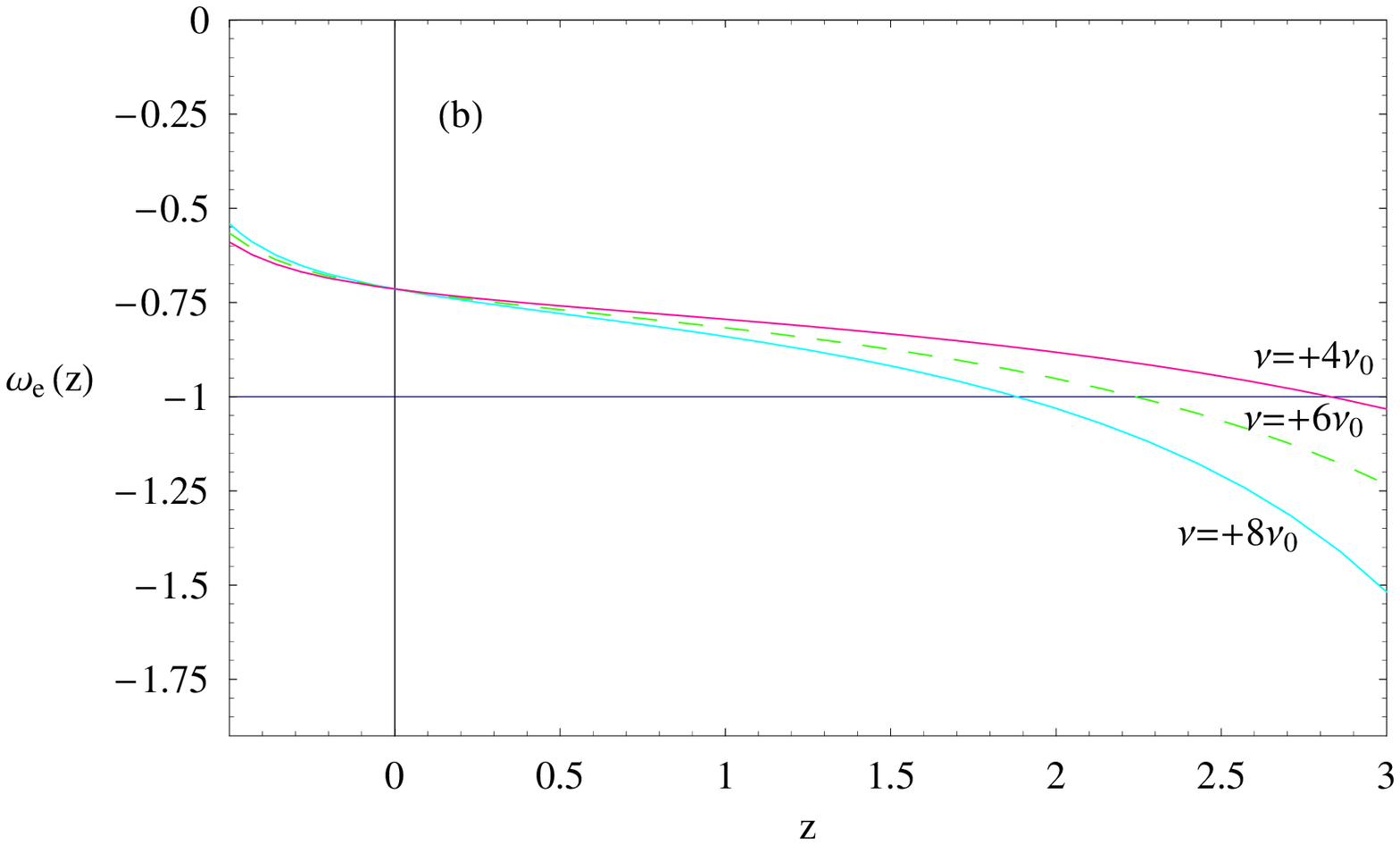}
\caption{The effective EOS (\protect\ref{eEOS}) of the $\Lambda$XCDM
model as a function of the redshift for a set of allowed values, see
Fig.\,\protect\ref{fig8}: (a) $\wX=-1.6$, $\OLo=0.80$ (phantom-like
cosmon) for three values of $\nu$; (b) $\wX=-0.8$, $\OLo=-0.3$
(quintessence-like cosmon)  for three positive values of $\nu$ (no
negative values allowed in this case) } \label{EOSexamples}
\end{figure}


We emphasize that this preference for the phantom matter character
of the cosmon entity does not necessarily translate to the overall
effective EOS of the $\CC$XCDM model. This has already been
discussed after Eq.\,(\ref{check}). The quintessence-like behavior
($d\OD/da<0$) or phantom-like behavior ($d\OD/da>0$) of the total DE
of the $\CC$XCDM  model depends only on the sign of the product
$(1+\wX)\,\OX$ in the regions where $\OD$ is positive. Let us
consider some examples, particularly those in
Fig.\,\ref{EOSexamples}a, which correspond to allowed values of the
parameters. Since, in this case, $\OXo=-0.1$ is small, from
Eq.\,(\ref{OXa}) we can see that for sufficiently large $z$ (small
$a$), the last term on the \textit{r.h.s.} is the dominant one
(using the fact that $\epsilon\ll 1$). Therefore, we have the
following rule of thumb: the total DE of the $\CC$XCDM model behaves
as quintessence ($\we\gtrsim-1$) or phantom ($\we\lesssim-1$), if
the sign of $\nu\,(1+\wX)/\wX$ is positive or negative,
respectively:
\begin{equation}\label{thumb}
1+\we(z)\gtrless 0\Longleftrightarrow
\frac{\nu\,(1+\wX)}{\wX}\gtrless 0\,.
\end{equation}
For example, if $X$ behaves as phantom matter (which is the
preferred state of the cosmon, according to our analysis), the
$\CC$XCDM model can be phantom-like if $\nu<0$, but behaves like
quintessence if $\nu>0$ (cf. Fig.\,\ref{EOSexamples}a). When the
first term on the \textit{r.h.s.} of Eq.\,(\ref{OXa}) becomes
important, the rule applies only if $z$ is sufficiently large, as it
is the case of the examples in Fig.\,\ref{EOSexamples}b, where
$\OXo=1$. These examples illustrate that the overall behavior of the
effective EOS of the $\CC$XCDM model is not dictated by the cosmon
EOS alone, but depends on other parameters, in particular $\nu$.

Finally, we may compare Fig.\,\ref{EOSexamples} of the $\CC$XCDM
model with Fig.\,\ref{weffigLtCDM} of the running $\CC$CDM model,
studied in the previous section. For the $\CC$XCDM, the allowed
values of $\nu$ are larger. As a result, the departure of the
effective EOS of this model with respect to the CC divide, $\we=-1$,
can be significantly greater in the redshift ranges that are
accessible to the next generation of supernovae experiments. It is,
therefore, possible to investigate deviations of $\we$ from -1,
predicted by the $\CC$XCDM model, in these experiments, particularly
by SNAP\,\cite{SNAP}. For instance, in Fig.\,\ref{EOSexamples}a we
see that for $\nu=-\nu_0$,  we have the predicted value
$\we(z=1.7)=-1.06$ and for $\nu=+\nu_0$, we have $\we(z=1.7)=-0.92$,
corresponding to the highest redshift accessible to SNAP. The
departure from $-1$ is significant large to be detected. At $z=2$,
which will be accessible in the future, we have $\we=-1.10$ and
$\we=-0.90$ for the same values of $\nu$.

\section{Conclusions\label{sect:conclusions}}

In this paper, we have explored the growth of matter density
perturbations for two running $\CC$ models in the literature, the
running $\CC$CDM model\,\cite{JHEPCC1,IRGA03,RGTypeIa} and the
$\CC$XCDM model\,\cite{LXCDM1,GSSIRGAC12}. These models offer
alternative explanations for the dynamical dark energy (DE), beyond
the usual proposals based on quintessence ideas. In particular, the
$\CC$XCDM model constitutes a promising cosmological framework, of a
very general nature, that also has the capacity to try to understand
the conspicuous cosmic coincidence problem, namely the perplexing
coincidence of finding ourselves, at present, in an expanding
Universe, where the amount of dark energy turns out to be of the
same order as that of matter. The cosmic coincidence is a riddle,
wrapped in the polyhedric mystery of the Cosmological Constant
Problem\,\cite{weinRMP}, which has many faces. In this case, the
conundrum is to understand the following situation. The density of
matter is continuously decreasing with the expansion of the
Universe, whereas the DE energy remains constant in the standard
$\CC$CDM model. It is totally inexplicable to understand from this
model, the fact that the cosmological constant line crosses the
ever-falling matter density curve, precisely now. The  $\CC$XCDM
model may be able to provide a clue to the resolution of this
enigma, due to the dynamical interplay of the cosmon $X$ and the
variable $\CC$, which together, form a composite dark energy medium.
This dynamics insures that the ratio of the DE to the matter density
stays bounded and that its value, at present, is not very different
from the value it will have in, say, the next Hubble time.

An important aspect of the $\CC$XCDM model that has been yet to be
investigated is its consistency with the LSS data. In this paper, we
have undertaken a thorough study, in an attempt to try to answer
this crucial question, namely, is there a non-empty region in the 3D
parameter space of the $\CC$XCDM model, capable of solving the
cosmic coincidence problem and, at the same time, being consistent
with the known data on structure formation?  Our study shows that
the answer to this question was, in fact, positive.

The ``effective approach'' that we have used here is based on three
essential ingredients: i) the use of the effective equation of state
(EOS) representation of cosmologies with variable cosmological
parameters\,\cite{SS1,SS2}; ii) the calculation of the growth of
matter density fluctuations, using the EOS of the
DE\,\cite{Ferreira97,Ma,LinderJenkins03}; and iii) the application
of the ``F-test''\,\cite{Ftest,vdecay} to compare the model with the
LSS data. This three-step  methodology turned out to be a
streamlined strategy. Even if it is not a perfect procedure to
estimate the restrictions that structure formation imposes on a
given model of dark energy, it is nevertheless an economical and
efficient method to encapsulate the essential findings of the
full-fledged approach. That this is so can be argued on the grounds
of the various existing cross-checks on the constraints that LSS
data impose e.g. on the running $\CC$CDM model, the first dynamical
$\CC$ model that we addressed in this study. Our ``effective
approach'' provides a noticeable consistency  with the complete
calculation of matter and DE perturbations presented for the running
$\CC$CDM model in Ref.\,\cite{FSS1}. We are, therefore, confident
that the same procedure is able to capture the main restrictions
that the present data put on the parameter space of the $\CC$XCDM
model.

In view of the consistency between the solution to the cosmic
coincidence problem proposed by the $\CC$XCDM model and the data on
structure formation, this model is substantially reinforced. It
emerges as a very strong candidate for a possible solution to the
cosmic coincidence problem. The physical region of its parameter
space turns out to be compatible with all cosmological data known at
present. Furthermore, we have shown (cf. Fig.\,\ref{EOSexamples})
that the model predicts non-trivial observable features in the EOS
of the dark energy. Most important, these features can be accessible
to the next generation of supernovae experiments, like DES and
SNAP\,\cite{SNAP}. This model, therefore, has the ability to solve
some of the important problems of modern cosmology and, in addition,
makes predictions which can be tested by observations from
experiments just around the corner. We eagerly await the possibility
of confronting the $\CC$XCDM model with these observations.

 {\bf Acknowledgments.} JG, AP and JS are partially supported by
MECYT and FEDER under project 2004-04582-C02-01, and also by DURSI
Generalitat de Catalunya under project 2005SGR00564. RO is partially
supported by the Brazilian Agency CNPq (Grant no.300699/2006-91).


\newcommand{\JHEP}[3]{ {JHEP} {#1} (#2)  {#3}}
\newcommand{\NPB}[3]{{\sl Nucl. Phys. } {\bf B#1} (#2)  {#3}}
\newcommand{\NPPS}[3]{{\sl Nucl. Phys. Proc. Supp. } {\bf #1} (#2)  {#3}}
\newcommand{\PRD}[3]{{\sl Phys. Rev. } {\bf D#1} (#2)   {#3}}
\newcommand{\PLB}[3]{{\sl Phys. Lett. } {\bf B#1} (#2)  {#3}}
\newcommand{\EPJ}[3]{{\sl Eur. Phys. J } {\bf C#1} (#2)  {#3}}
\newcommand{\PR}[3]{{\sl Phys. Rep } {\bf #1} (#2)  {#3}}
\newcommand{\RMP}[3]{{\sl Rev. Mod. Phys. } {\bf #1} (#2)  {#3}}
\newcommand{\IJMP}[3]{{\sl Int. J. of Mod. Phys. } {\bf #1} (#2)  {#3}}
\newcommand{\PRL}[3]{{\sl Phys. Rev. Lett. } {\bf #1} (#2) {#3}}
\newcommand{\ZFP}[3]{{\sl Zeitsch. f. Physik } {\bf C#1} (#2)  {#3}}
\newcommand{\MPLA}[3]{{\sl Mod. Phys. Lett. } {\bf A#1} (#2) {#3}}
\newcommand{\JPA}[3]{{\sl J. Phys.} {\bf A#1} (#2) {#3}}
\newcommand{\CQG}[3]{{\sl Class. Quant. Grav. } {\bf #1} (#2) {#3}}
\newcommand{\JCAP}[3]{{ JCAP} {\bf#1} (#2)  {#3}}
\newcommand{\APJ}[3]{{\sl Astrophys. J. } {\bf #1} (#2)  {#3}}
\newcommand{\AMJ}[3]{{\sl Astronom. J. } {\bf #1} (#2)  {#3}}
\newcommand{\APP}[3]{{\sl Astropart. Phys. } {\bf #1} (#2)  {#3}}
\newcommand{\AAP}[3]{{\sl Astron. Astrophys. } {\bf #1} (#2)  {#3}}
\newcommand{\MNRAS}[3]{{\sl Mon. Not. Roy. Astron. Soc.} {\bf #1} (#2)  {#3}}



\end{document}